\documentclass[%
 reprint,
 amsmath,amssymb,
 aps,
]{revtex4-2}

\usepackage{graphicx}
\usepackage{dcolumn}
\usepackage{bm}
\usepackage{hyperref}
\usepackage{upgreek}
\usepackage{soul,color}

\usepackage{xcolor}

\begin{document}


\title{Exact electromagnetic multipole expansion \\using elementary current multipoles}

 \author{Radoslaw Kolkowski}
 \email{radoslaw.kolkowski@aalto.fi}
 \author{Sagar Sehrawat}%
 \email{sagar.sehrawat@aalto.fi}
\author{Andriy Shevchenko}
\email{andriy.shevchenko@aalto.fi}
\affiliation{%
Department of Applied Physics, Aalto University, P.O.Box 13500, Aalto FI-00076, Finland
}%

\begin{abstract}
Multipole expansion plays an important role in the description of electromagnetic scatterers, allowing them to be accurately characterized by a small set of expansion coefficients. However, to describe electromagnetic excitations inside a scatterer, the current density in it should be decomposed into \emph{current multipoles}, which include nonradiating current configurations (anapoles) that are absent in the classical field-based expansion. Unfortunately, the use of current multipoles has so far been limited by the absence of an exact and general expression for the current multipole moments beyond their point-multipole approximation. Here, we derive such an expression and present the exact mapping relations between the classical and current multipole moments. We use our theory to calculate the scattering and extinction cross sections for large, wavelength-scale, optical scatterers supporting multipole excitations up to the sixth order, showing perfect agreement with the Mie theory. 
We also demonstrate the ability of current multipole expansion to describe anapole excitations beyond the small-scatterer approximation, which allows us to derive the \emph{exact anapole condition} and reveal the actual current configurations and their contributions to scattering. Our theoretical framework is valid for electromagnetic scatterers of arbitrary sizes and shapes without restrictions on the multipole orders, complementing the existing theory of electromagnetic multipole expansion. The minimalistic and universal character of current multipoles makes them a convenient tool for characterizing and designing diverse electromagnetic scattering systems of arbitrary complexity.

\end{abstract}

\maketitle

\section{Introduction}

The radiation and scattering of electromagnetic fields are among the most fundamental phenomena in physics. Their rigorous theoretical treatment is essential in many areas of science and technology, from telecommunications~\cite{balanis2016} to atmospheric science~\cite{rayleigh1899,bohren2008} and astrophysics~\cite{huffman1977}. Recently, scattering theory has become particularly relevant for the rapidly growing field of nanophotonics, in which the resonant optical response of nanostructured media has created new paradigms for optical field manipulation and enhanced light-matter interactions~\cite{koenderink15,schulz24}. For example, multipole excitations at Mie resonances~\cite{kivshar2017,kruk2017,koshelev2020,liu2020, babicheva2024} have been extensively explored for the realization of Kerker effect~\cite{butakov2016,liu2018,shamkhi2019}, toroidal and anapole excitations~\cite{liu2017_,tuz20,canos2021,bukharin26}, bound states in the continuum~\cite{koshelev2019,sadrieva2019,chen2019}, antireflection coatings~\cite{spinelli2012}, Huygens metasurfaces~\cite{decker2015,liu2017}, nanoscale lasers~\cite{tiguntseva2020}, metamaterials~\cite{petschulat2008,grahn2013}, structural colors~\cite{baek2020}, nonlocal (collective) resonances~\cite{evlyukhin2012,kwon2018,kolkowski2023,kolkowski2024}, and enhanced nonlinear optical effects~\cite{smirnova2016,sain2019}.

Theoretical description and design of the scattering properties of micro- and nanoscale objects can be done using electromagnetic multipole expansion. It allows one to characterize every finite-sized scattering system with a relatively small set of expansion coefficients corresponding to mutually orthogonal electromagnetic fields of different-order multipoles~\cite{jackson2012}. Since the original Mie theory~\cite{mie1908}, multipole expansion has been approached in many ways~\cite{muhlig2011,evlyukhin2011,grahn2012,evlyukhin2013,evlyukhin2016,alaee2018,alaee2019,evlyukhin2019,gurvitz2019,majorel2022,ospanova23,evlyukhin23}. However, regardless of the approach, the classical field-based expansion does not fully capture the actual current oscillations within the system. For example, electromagnetic anapoles are absent in this expansion, as they do not radiate into the far field~\cite{savinov2019,baryshnikova2019,koshelev2019_,shevchenko2020}. As a result, theoretical description of anapoles often involves the so-called toroidal moments, which are artificially separated parts of the actual multipole moments~\cite{miroshnichenko2015,gurvitz2019,babicheva2024}. Since the conventional multipole expansion is based on the radiated fields rather than their sources (i.e., oscillating currents), it is difficult to use it for engineering the electromagnetic scatterers.

\begin{figure}[t]
\centering\includegraphics[width=8.5cm]{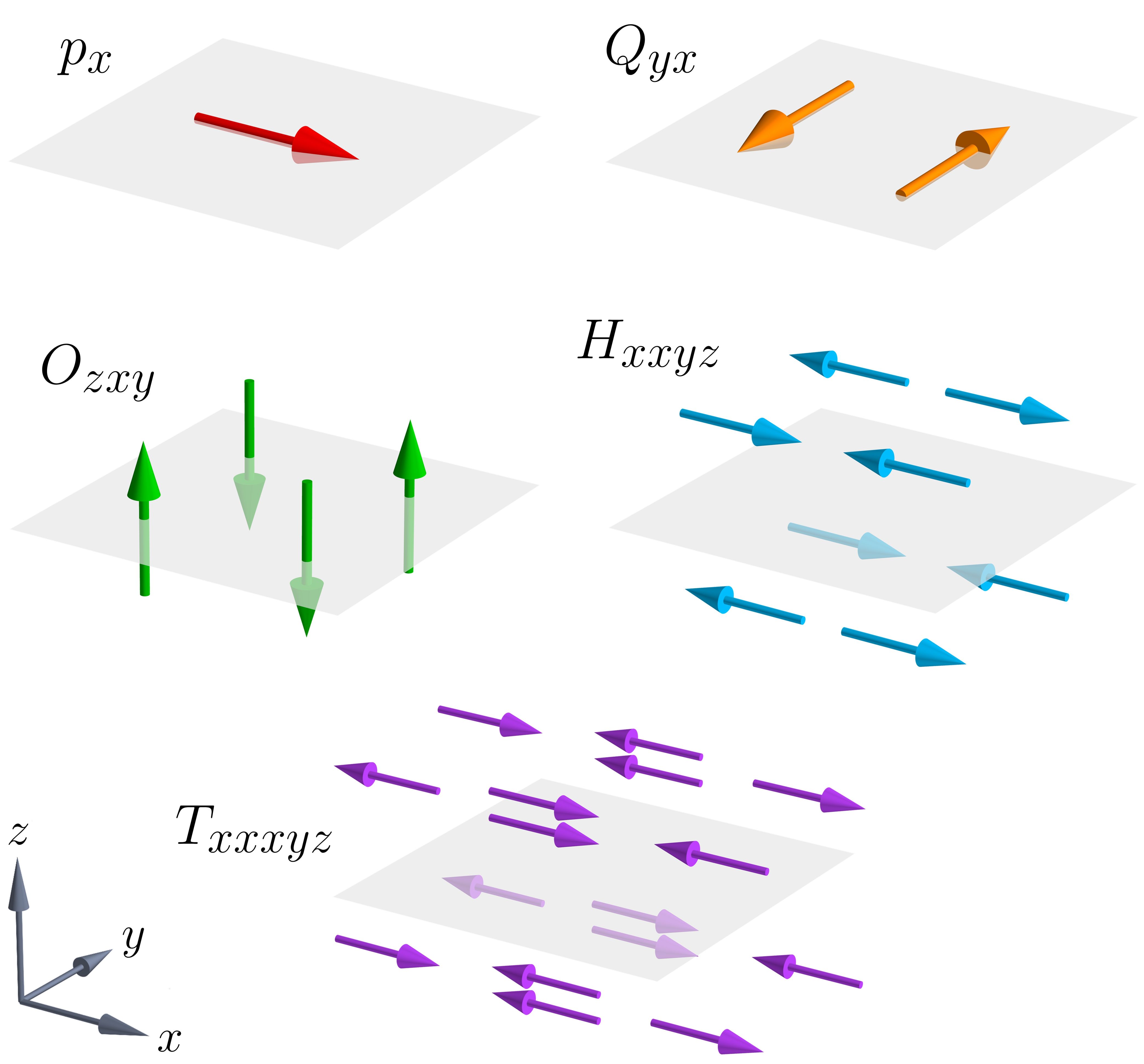}
\caption{\label{fig:1} Examples of current configurations in the excitations of current multipoles of different orders $l$ in the point-multipole approximation: a current dipole ($l$ = 1), quadrupole ($l$ = 2), octupole ($l$ = 3), hexadecapole ($l$ = 4), and triacontadipole ($l$ = 5), corresponding to moments $p_x$, $Q_{yx}$, $O_{zxy}$, $H_{xxyz}$, and $T_{xxxyz}$, respectively. Each arrow represents a single linear current element (see Ref.~\cite{grahn2012}). Current multipoles of high orders, such as 4 and 5, are relatively common, constituting classical multipoles of lower orders. In particular, current hexadecapoles are constituents of classical electric quadrupoles and magnetic octupoles, whereas current triacontadipoles appear in classical electric octupoles and magnetic hexadecapoles (see Eqs. (\ref{eq:aE20})-(\ref{eq:aM44})). In numerical calculations presented in this work, we consider classical multipoles up to the electric and magnetic hexacontatetrapoles ($l$ = 6)  which include current multipoles of orders $l$ up to 8. Note that, in the used convention, the outermost current element at positive coordinates is positive if the multipole moment is positive.}
\end{figure}

To overcome these limitations, a complementary multipole expansion has been proposed~\cite{grahn2012}. It expands the scattering current density $\textbf{J}(\textbf{r})$ in the scatterer, defined as
\begin{equation}
\label{eq:J}
\textbf{J}(\textbf{r})=-i\omega\epsilon_0[\epsilon_{\text{r}}(\mathbf{r})-\epsilon_{\text{s}}]\textbf{E}(\textbf{r}),
\end{equation}
in terms of the current multipoles. In the point-multipole approximation, the moments of these multipoles are introduced by the following multipole tensor:
\begin{equation}
\label{eq:M_point}
\boldsymbol{M}^{(l)}_{\text{point}}=\frac{i}{(l-1)!\,\omega}\int\limits^{\infty}_{-\infty} \textbf{J}(\textbf{r})\textbf{r}^{l-1}d^3\textbf{r},
\end{equation}
similarly to the moments of the primitive Cartesian multipoles~\cite{harrington1961}. In the above equations, $\epsilon_{\text{r}}(\mathbf{r})$ is the position-dependent relative electric permittivity and $\epsilon_{\text{s}}$ is its value in the surrounding medium, $l$ is the multipole order, and $\mathbf{J}(\mathbf{r})\textbf{r}^{l-1}$ is the outer product of vector $\mathbf{J}(\mathbf{r})$ and $l-1$ vectors $\mathbf{r}$. The current multipoles are not divided into electric and magnetic multipoles and represent very simple configurations of linear current elements (see Fig.~\ref{fig:1}), which facilitates their use in designing individual and arrayed scatterers with prescribed near-field characteristics~\cite{grahn2012_,shevchenko2015,shevchenko2020,sehrawat2024,sehrawat2024_,sehrawat2024__}.

\newpage

\begin{table}[t]
\caption{\label{tab:table1}%
Exact expressions for the current multipole moments illustrated schematically in Fig.~\ref{fig:1}. The names of the multipoles follow the same convention as in Refs.~\cite{frizyuk2019,palazzo2023}.
}
\begin{ruledtabular}
\begin{tabular}{llcc}
$l$ &\textrm{Name}&\textrm{Symbol}&\textrm{Expression}\\
\colrule\vspace{-2.5mm}\\ 
1 & Dipole & $\displaystyle p_{x} $ & $\displaystyle\frac{i}{\omega}\int\limits_{-\infty}^{\infty} j_{0}\left(k r\right) J_{x}(\textbf{r}) \,d^3\textbf{r}$ \\ \vspace{-2.5mm}\\
2 & Quadrupole & $\displaystyle Q_{yx} $ & $\displaystyle\frac{3 i }{\omega k}\int\limits_{-\infty}^{\infty} j_{1}\left(k r\right)J_{y}(\textbf{r}) \frac{x}{r} d^3\textbf{r}$\\ \vspace{-2.5mm}\\
3 & Octupole & $\displaystyle O_{zxy} $ & $\displaystyle\frac{15 i }{2\omega k^2}\int\limits_{-\infty}^{\infty} j_{2}\left(k r\right)J_{z}(\textbf{r}) \frac{xy}{r^2}d^3\textbf{r}$ \\ \vspace{-2.5mm}\\
4 & Hexadecapole & $\displaystyle H_{xxyz} $ & $\displaystyle\frac{35 i }{2\omega k^3}\int\limits_{-\infty}^{\infty} j_{3}\left(k r\right)J_{x}(\textbf{r}) \frac{xyz}{r^3} d^3\textbf{r}$ \\ \vspace{-2.5mm}\\
5 & Triacontadipole & $\displaystyle T_{xxxyz} $ & $\displaystyle\frac{315 i }{8\omega k^4}\int\limits_{-\infty}^{\infty} j_{4}\left(k r\right)J_{x}(\textbf{r}) \frac{x^2 yz}{r^4} d^3\textbf{r}$ \\ \vspace{-2.5mm}\\
\end{tabular}
\end{ruledtabular}
\end{table}

Unfortunately, Eq. (\ref{eq:M_point}) is not valid beyond the long-wavelength (point-multipole) approximation and cannot be used to accurately evaluate the magnitudes of the multipole moments and their contributions to the scattering and extinction cross sections for wavelength-scale and larger scatterers. So far, the only method for obtaining the exact current multipole moments for such scatterers was to calculate them from the exact moments of the classical electric and magnetic multipoles, i.e., from the multipole expansion coefficients $a_{\text{E}}(l, m)$ and $a_{\text{M}}(l,m)$, where $l$ and $m$ are the expansion orders~\cite{grahn2012,grahn2013,sehrawat2024,sehrawat2024_,sehrawat2024__}. However, because current multipoles include nonradiating current configurations that are absent in the classical multipole expansion, the mapping from the classical to current multipole moments is underdetermined, so that some of the current multipole moments cannot be unambiguously evaluated~\cite{sehrawat2024}. The lack of a complete theoretical framework for the current multipole expansion has so far prevented researchers from using the approach to its full potential, despite its clear advantages that make it complementary to the classical multipole expansion.

\begin{figure*}[t]
\centering\includegraphics[width=17cm]{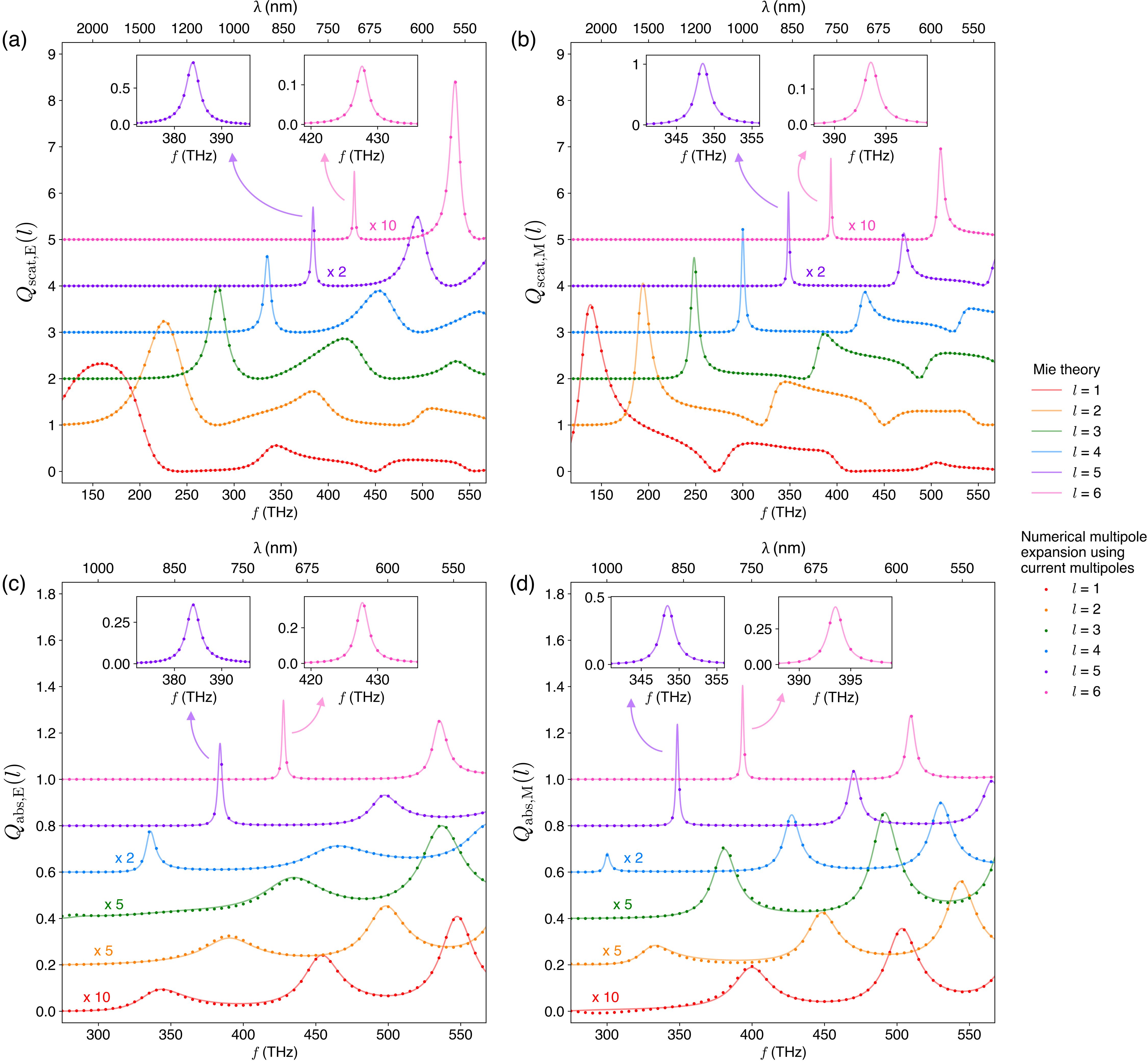}
\caption{\label{fig:2} Numerical validation of the exact expression for the current multipole moments [Eq.~(\ref{eq:M_exact})] using the optical response of a silicon sphere of diameter 600 nm embedded in PMMA, for which we evaluate the multipole contributions to the normalized scattering cross section $Q_{\text{scat}}$ (a, b) and absorption cross section $Q_{\text{abs}} = Q_{\text{ext}}-Q_{\text{scat}}$ (c, d) defined in Eqs.~(\ref{eq:scat})-(\ref{eq:abs}). The dots correspond to the contributions of moments $a_{\text{E}}(l,m)$ and $a_{\text{M}}(l,m)$ obtained by numerically calculating the current multipole moments  $M^{(l)}_{\text{exact}}$ in \textsc{Comsol Multiphysics} using Eqs.~(\ref{eq:J}) and (\ref{eq:M_exact}) and converting them to the classical multipole moments using Eqs.~(\ref{eq:Mapping-for-aE0})-(\ref{eq:Mapping-for-aM1}), (\ref{eq:aE20})-(\ref{eq:aM44}), and Ref.~\cite{kolkowski2025}. The solid lines are the cross sections obtained from the Mie theory. The insets show the peaks of electric and magnetic triacontadipoles ($l=5$) and hexacontatetrapoles ($l=6$). In the main plots, the curves for subsequent multipoles are vertically offset for clarity, and some of them are magnified by a constant factor, as indicated. The results of numerical multipole expansion and Mie theory overlap perfectly (up to the numerical precision of \textsc{Comsol}), which validates the expressions presented in this work.}
\end{figure*}

Here, we derive the long-sought general expression for the current multipole moments of any order $l$. The expression is valid for scatterers of arbitrary sizes and shapes. The obtained general current-multipole tensor reads:
\begin{equation}
\label{eq:M_exact}
\boldsymbol{M}^{(l)}_{\text{exact}}=\frac{i}{\omega}\frac{(2l-1)!!}{(l-1)!}\int\limits_{-\infty}^{\infty} \textbf{J}(\textbf{r})\textbf{r}^{l-1}\frac{j_{l-1}(kr)}{(kr)^{l-1}}d^3\textbf{r},
\end{equation}
where $(2l-1)!!=(2l-1)\cdot(2l-3)\cdot...\cdot5\cdot3\cdot1$ is the double factorial, $j_{l-1}(kr)$ are the spherical Bessel functions of the first kind, and $k$ is the wavenumber in the surrounding medium. The derivation of this expression is presented in Appendix A. Equation (\ref{eq:M_exact}) allows the moments, i.e., the elements of the tensor $\boldsymbol{M}^{(l)}_{\text{exact}}$, to be calculated directly and unambiguously from the scattering current density distribution $\mathbf{J}(\mathbf{r})$ defined in Eq. (\ref{eq:J}). Moreover, the moments can be used to assess the magnitude of the corresponding current density (see Appendix B and Refs.~\cite{sehrawat2024_,sehrawat2024__}). Several examples of these moments are presented in Table \ref{tab:table1}, and the corresponding configurations of linear current elements in the point-multipole approximation are shown schematically in Fig.~\ref{fig:1}. The expressions in Table~\ref{tab:table1} illustrate the fact that every current multipole moment corresponds to a unique monomial containing a Cartesian vector component of the scattering current density ($J_x$, $J_y$, $J_z$) and, for $l\geq2$, a product of normalized Cartesian coordinates $x/r$, $y/r$, and $z/r$. These moments constitute exceptionally simple Cartesian configurations of oscillating currents, which makes them truly the \emph{elementary} multipole moments. At the same time, they are exact and applicable to scatterers of arbitrary sizes and shapes, which we demonstrate in the next section.

\begin{figure*}[t]
\centering\includegraphics[width=17cm]{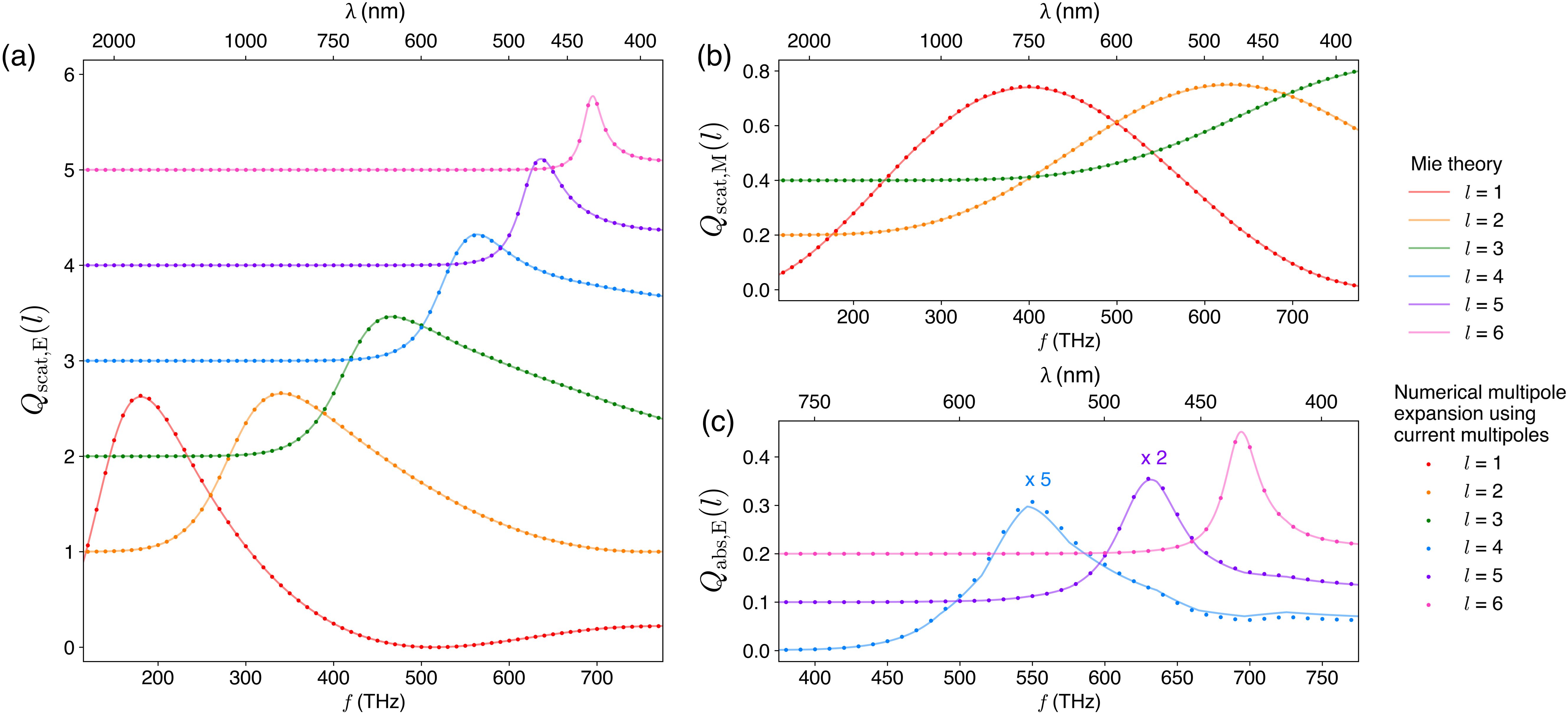}
\caption{\label{fig:3} Numerical validation of the current multipole expansion using the optical response of a silver sphere of diameter 400 nm embedded in PMMA. Similarly to Fig.~\ref{fig:2}, the solid lines represent the cross sections calculated using the Mie theory, while the dots correspond to the cross sections obtained numerically in \textsc{Comsol Multiphysics} using the current multipole expansion (Eqs.~(\ref{eq:J}), (\ref{eq:M_exact}), (\ref{eq:Mapping-for-aE0})-(\ref{eq:Mapping-for-aM1}), (\ref{eq:aE20})-(\ref{eq:aM44}), and Ref.~\cite{kolkowski2025}). In the given example, the scattering cross section ($Q_{\text{scat}}$) has significant contributions of (a) the electric multipoles of order $l\leq6$ (up to hexacontatetrapoles) and (b) the magnetic multipoles of order $l\leq3$ (up to octupoles), while the absorption cross section ($Q_{\text{abs}}$) is dominated by the electric multipoles of $l\geq$4, as shown in (c). Similarly to Fig.~\ref{fig:2}, we find an excellent agreement between the numerical multipole expansion and Mie theory. }
\end{figure*}

\section{Numerical validation}

To validate Eq.~(\ref{eq:M_exact}), we used it to numerically calculate the multipole contributions to the scattering and absorption cross sections for wavelength-scale scatterers across the visible and near-infrared spectral range. We did this by first calculating the elements of tensor $\boldsymbol{M}^{(l)}_{\text{exact}}$, and then using them to calculate the multipole expansion coefficients $a_{\text{E}}(l, m)$ and $a_{\text{M}}(l, m)$. Finally, we evaluated their contributions to the cross sections in question. 

The exact expressions for coefficients $a_{\text{E}}(l, m)$ and $a_{\text{M}}(l, m)$ in terms of the generalized current multipole moments are derived in Ref.~\cite{kolkowski2025}. The moments of the two expansions are just simple linear combinations of each other. For example, the classical electric and magnetic dipole moments are written as
\begin{eqnarray}
a_\text{E}(1,0) =&&\, \sqrt{2}C_{1} p_{z}+\frac{7\sqrt{2}}{3} C_{3}(3 O_{xxz}+3 O_{yyz} \nonumber \\&& 
- O_{zxx} - O_{zyy} + 2 O_{zzz}), \label{eq:Mapping-for-aE0} 
\end{eqnarray}
\begin{eqnarray}
a_\text{E}(1,\pm 1) =&&\, C_{1} (\mp p_{x} + i p_{y}) \mp \frac{7}{3} C_{3} [2 O_{xxx}- O_{xyy} \nonumber \\
&& - O_{xzz} + 3 O_{yxy} + 3 O_{zxz} \mp i (3 O_{xxy} \nonumber \\
&&- O_{yxx} + 2 O_{yyy} - O_{yzz} + 3 O_{zyz})],\label{eq:Mapping-for-aE1} 
\end{eqnarray}
\begin{eqnarray}
a_\text{M}(1,0)=&&\, 5 \sqrt{2} i C_{2} (- Q_{xy} + Q_{yx}),\label{eq:Mapping-for-aM0} 
\end{eqnarray}
\begin{eqnarray}
a_\text{M}(1,\pm1)\!=&&\, 5 C_{2} [- Q_{xz} + Q_{zx} \pm i (Q_{yz} - Q_{zy})],\label{eq:Mapping-for-aM1} 
\end{eqnarray}
where $C_{1} = - i k^{3}/(6 \pi\epsilon E_{0})$, $C_{2} = - k^{4}/(60 \pi\epsilon E_{0})$, and $C_{3} = -i k^{5}/(210 \pi\epsilon E_{0})$, with $E_{0}$ being the incident field amplitude and $\epsilon=\epsilon_{0}\epsilon_{\text{s}}$. The mapping relations for a set of higher-order multipoles are presented in Appendix C. Additionally, in Ref.~\cite{kolkowski2025} we provide a Python code to automatically derive the mapping relations for an arbitrary order $l$.

After obtaining $a_\text{E}(l,m)$ and $a_\text{M}(l,m)$, the multipole contributions to the normalized scattering and extinction cross sections at each order $l$ are given by
\begin{equation} \label{eq:scat}
Q_{\text{scat,E/M}}(l)=\frac{\pi}{k^2 S}\sum_{m=-l}^{l}(2l+1)|a_\text{E/M}(l,m)|^2,    
\end{equation}
\begin{equation} \label{eq:extE}
Q_{\text{ext,E}}(l)=-\frac{\pi}{k^2 S}\sum_{m=-1,+1}(2l+1)\operatorname{Re}[ma_\text{E}(l,m)],
\end{equation}
\begin{equation} \label{eq:extM}
Q_{\text{ext,M}}(l)=-\frac{\pi}{k^2 S}\sum_{m=-1,+1}(2l+1)\operatorname{Re}[a_\text{M}(l,m)],
\end{equation}
where $S$ is the geometric cross section of the scatterer ($S = \pi R^2$ for a spherical nanoparticle of radius $R$). 
The corresponding multipole contributions to the normalized absorption cross section are then calculated as
\begin{equation} \label{eq:abs}
Q_{\text{abs,E/M}}(l)=Q_{\text{ext,E/M}}(l)-Q_{\text{scat,E/M}}(l).
\end{equation}

In the numerical calculations, $\mathbf{J}(\mathbf{r})$ defined in Eq.~(\ref{eq:J}) was obtained from the electric field distribution calculated in \textsc{Comsol Multiphysics} for a silicon sphere of diameter 600 nm (see Fig.~\ref{fig:2}) and a silver sphere of diameter 400 nm (see Fig.~\ref{fig:3}). In both cases, we assumed the scatterers to be embedded in PMMA, which is a dielectric material commonly used in nanofabrication. The optical constants of Si, Ag, and PMMA were taken from Refs.~\cite{polyanskiy2024,franta2017,johnson1972,beadie2015}. The scatterers were illuminated by a plane wave linearly polarized along the $x$ axis and propagating along the $z$ axis, which corresponds to the definition of the extinction cross section given in Eqs.~(\ref{eq:extE}) and (\ref{eq:extM}). In the calculation of multipole moments, the coordinate system had its origin at the center of the scatterer. Figures \ref{fig:2} and \ref{fig:3} show the calculated cross sections. In the presented plots, the dots were obtained using the current multipole moments computed numerically in \textsc{Comsol Multiphysics} with the aid of Eq.~(\ref{eq:M_exact}) together with Eqs.~(\ref{eq:J}), (\ref{eq:Mapping-for-aE0})-(\ref{eq:abs}), (\ref{eq:aE20})-(\ref{eq:aM44}) and Ref.~\cite{kolkowski2025}, while the solid lines were calculated directly using the Mie theory. The results obtained with both approaches are seen to be in excellent agreement. 

In the considered examples, the scatterers are comparable in size to the wavelength and cannot be treated as point scatterers. Using the multipole expansion in the long-wavelength (small-scatterer) approximation for such scatterers is known to produce significant errors~\cite{alaee2018}. Hence, the presented analysis of wavelength-scale scatterers confirms the validity of our theory beyond the long-wavelength approximation, demonstrating its applicability to scatterers of an arbitrary size.

It is illuminating to note that Eqs.~(\ref{eq:Mapping-for-aE0}) and (\ref{eq:Mapping-for-aE1}) for the classical electric dipole moments contain the current dipole and octupole moments, while Eqs.~(\ref{eq:Mapping-for-aM0}) and (\ref{eq:Mapping-for-aM1}) for the classical magnetic dipole moments contain the current quadrupole moments. In fact, every classical electric multipole moment of order $l$ can be expressed as a superposition of current multipole moments of orders $l$ and $l+2$, while every classical magnetic multipole moment can be written as a superposition of current multipole moments of order $l+1$. This shows that the magnetic dipoles (or, more generally, magnetic multipoles of order $l$) in the field-based expansion are in fact excitations of the same order as the electric quadrupoles (or, more generally, multipoles of order $l+1$) in the current expansion. Moreover, the fact that every classical electric multipole moment of order $l$ contains a more complex current excitation of order $l+2$ is partially associated with the emergence of the so-called toroidal moments. The relevance of these findings for anapole excitations is discussed in the next section.

\section{Anapoles}

\begin{figure}[t]
\centering\includegraphics[width=8.5cm]{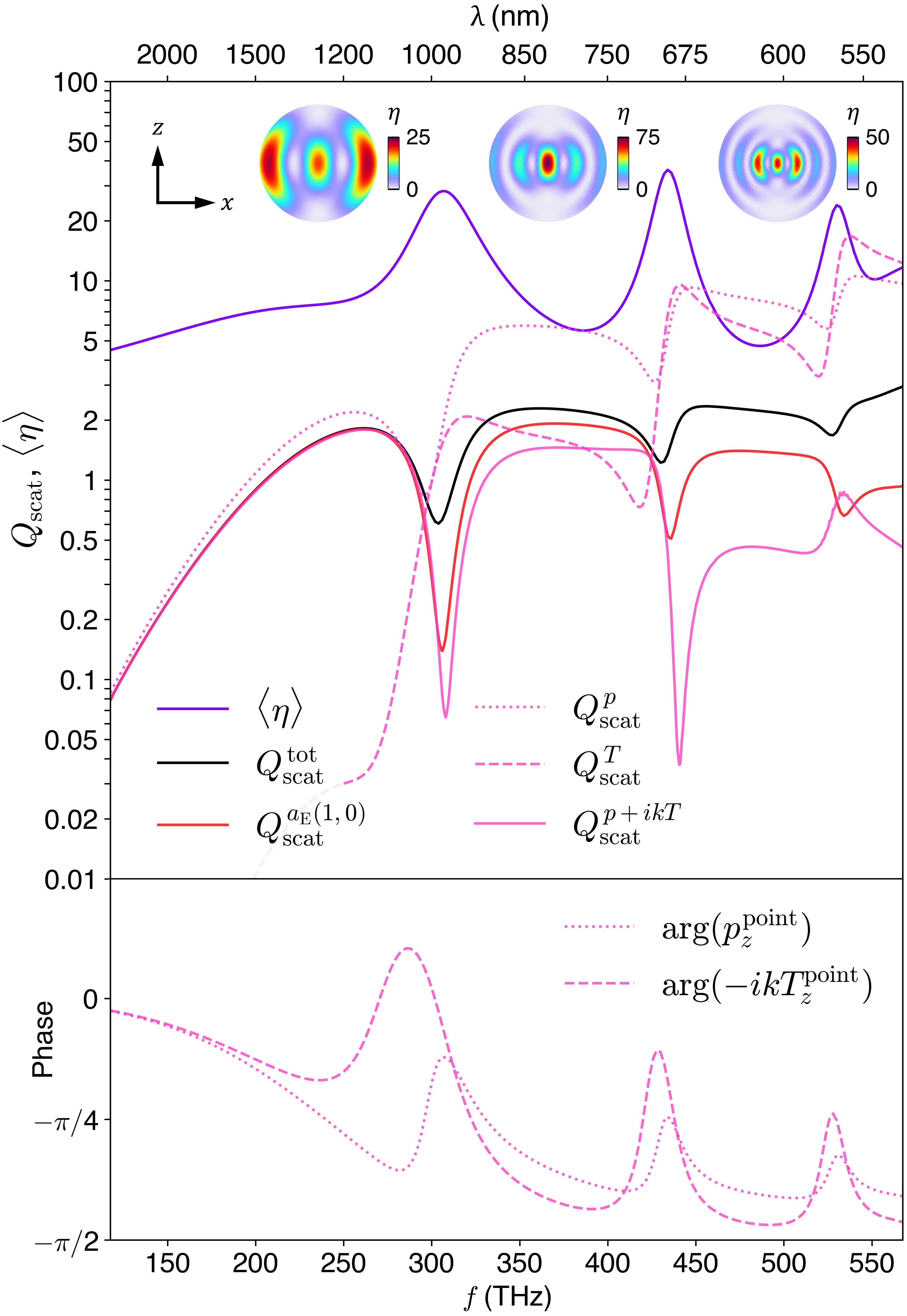}
\caption{\label{fig:4} Anapole excitations in a silicon nanodisk: conventional analysis based on decomposition of the total electric dipole moment into the electric dipole and electric toroidal dipole moments in the point-scatterer approximation. In this example, we consider a disk of diameter 600 nm and thickness 60 nm, surrounded by vacuum, illuminated by a linearly polarized plane wave. The anapole excitations are revealed by peaks in the spectrum of volume-averaged electric energy density enhancement $\langle\eta\rangle$ (violet curve), coinciding with the dips of the total scattering cross section $Q_{\text{scat}}^{\text{tot}}$ (black) and the electric-dipole scattering cross section $Q_{\text{scat}}^{a_\text{E}(1,0)}$ (red). The pink curves in the upper plot show the contributions of $p_z^{\text{point}}$, $T_z^{\text{point}}$ and $p_z^{\text{point}}+ikT_z^{\text{point}}$ to the scatteting cross section (dotted, dashed, and solid curve, respectively; see Eqs.~(\ref{eq:pzsph})-(\ref{eq:Tz})), whereas the phases of $p_z^{\text{point}}$ and $-ikT_z^{\text{point}}$ are shown in the lower plot (dotted and dashed, respectively). The insets show the spatial distributions of $\eta$ across the $xz$-plane inside the disk for each of the anapole excitations (at $f$ = 307, 434, and 530 THz).}
\end{figure}

In this section, we show that the current multipole expansion not only enables a complementary understanding of the anapole excitations but also can be used to refine and generalize the anapole condition. Let us start with the well-known example of anapole excitations inside a dielectric nanodisk ~\cite{miroshnichenko2015,zenin2017}. We assume that the disk is made of silicon and surrounded by vacuum. The diameter of the disk (in the $xz$-plane) is 600 nm, and the thickness (along the $y$-axis) is 60 nm. The disk is illuminated by a plane wave, linearly polarized along the $z$-axis and propagating along the $y$-axis. Just like in the previous section, we study this system numerically using \textsc{Comsol Multiphysics}. 

The violet curve in Fig.~\ref{fig:4} shows the calculated spectral dependence of the enhancement of the volume-averaged electric energy density, $\langle\eta\rangle = |\epsilon_r| \,\langle|\mathbf{E}|^2\rangle/E_0^2$. We plot it together with the total scattering cross section $Q_{\text{scat}}^{\text{tot}}$ (black curve), which is obtained by integrating the radial component of the Poynting vector of the scattered field over the surface of a sphere enclosing the disk. In addition to $Q_{\text{scat}}^{\text{tot}}$, the red curve in Fig.~\ref{fig:4} shows the contribution of the electric-dipole scattering, which is in this case dominated by the expansion coefficient $a_{\text{E}}(1,0)$. The spectrum of $\langle\eta\rangle$ has three pronounced peaks that coincide with the dips in both $Q_{\text{scat}}^{\text{tot}}$ and $Q_{\text{scat}}^{a_{\text{E}}(1,0)}$, indicating the excitation of electric-dipole anapoles suppressing the electric-dipole scattering with the dipole moment oriented along the $z$-axis. The insets in Fig.~\ref{fig:4} reveal the electric field distributions at the peaks of $\langle\eta\rangle$, allowing us to identify the excitations as fundamental ($f$ = 307 THz) and higher-order anapole excitations ($f$ = 434 and 530 THz). 

Such electric-dipole anapoles are typically analyzed by expressing the electric dipole moment deduced from the far-field ($\mathbf{p}^{\text{far-field}}$) as a sum of the electric dipole and toroidal electric dipole moments in the point-scatterer approximation ($\mathbf{p}^{\text{point}}$ and $\mathbf{T}^{\text{point}}$). For a dipole moment oriented along the $z$-axis, the amplitude can be expressed as
\begin{equation} \label{eq:pzsph}
p_z^{\text{far-field}}=\frac{1}{\sqrt{2}C_1}a^{\text{point}}_{\text{E}}(1,0)=p_z^{\text{point}}+ikT_z^{\text{point}},    
\end{equation}
where
\begin{equation} \label{eq:pz-small}
p_z^{\text{point}}=\frac{i}{\omega}\int\limits_{-\infty}^{\infty}J_z(\mathbf{r})d^3\mathbf{r}    
\end{equation}
and the toroidal electric dipole moment is
\begin{equation} \label{eq:Tz}
T_z^{\text{point}}=\frac{k}{10\omega}\int\limits_{-\infty}^{\infty}\left\{[\mathbf{r}\cdot\mathbf{J}(\mathbf{r})]z-2r^2J_z(\mathbf{r})\right\}d^3\mathbf{r}.    
\end{equation}
Note that $p_z^{\text{point}}$ does not depend on the choice of the origin of the coordinate system, as opposed to $T_z^{\text{point}}$. For example, shifting the origin to increase $r$ allows $T_z^{\text{point}}$ to become arbitrarily large. By convention, the origin is chosen at the geometric center of the disk, in which case the anapole condition is written as
\begin{equation}p_z^{\text{point}}+ikT_z^{\text{point}}=0.
\end{equation} 
Under this condition, the electric-dipole scattering due to $a_{\text{E}}^{\text{point}}(1,0)$ disappears. Using Eqs.~(\ref{eq:pzsph})-(\ref{eq:Tz}), this condition can be written in terms of the point current multipole moments, defined in Eq. (\ref{eq:M_point}), as  
\begin{equation} \label{eq:an-con-small}
p_z^{\text{point}}+\frac{k^2}{15} O_{a_{\text{E}}(1,0)}^{\text{point}}=0,
\end{equation}
where
\begin{equation} \label{eq:OaE10_point}
\begin{aligned}
O_{a_{\text{E}}(1,0)}^{\text{point}} = \frac{15i}{k}T_z^{\text{point}}=3O_{xxz}^{\text{point}}+3O_{yyz}^{\text{point}}\qquad\\-3O_{zzz}^{\text{point}}-6O_{zxx}^{\text{point}}-6O_{zyy}^{\text{point}}.
\end{aligned}
\end{equation}
This result is in agreement with Eq.~(43) of Ref.~\cite{grahn2012} after setting $a_{\text{E}}(1,0)$ to 0. It shows that, in the point-scatterer approximation, the toroidal dipole moment is equivalent to a sum of current octupole moments. However, scatterers supporting anapole excitations are typically large, at least on the order of $\lambda/2$~\cite{miroshnichenko2015,zenin2017,baryshnikova2019,savinov2019,canos2021}. Because of that, the point-scatterer approximation breaks down for higher-order anapoles. To illustrate this, let us consider the contributions of $p_z^{\text{point}}$ and $T_z^{\text{point}}$ to the scattering cross section:
\begin{equation}
Q_{\text{scat}}^{p}=\frac{3\pi}{k^2 S}\left|\sqrt{2}C_1 p_z^{\text{point}}\right|^2, 
\end{equation}
\begin{equation}
Q_{\text{scat}}^{T}=\frac{3\pi}{k^2 S}\left|\sqrt{2}C_1 ikT_z^{\text{point}}\right|^2, 
\end{equation}
\begin{equation}
Q_{\text{scat}}^{p+ikT}=\frac{3\pi}{k^2 S}\left|\sqrt{2}C_1 \left(p_z^{\text{point}}+ikT_z^{\text{point}}\right)\right|^2. 
\end{equation}
These quantities are shown as dotted, dashed, and solid pink curves in the upper part of Fig.~\ref{fig:4}, respectively. Anapoles are formed at frequencies at which $Q_{\text{scat}}^p = Q_{\text{scat}}^T$ and $p_z^{\text{point}}$ is out of phase with $ikT_z^{\text{point}}$ (see the lower part of Fig.~\ref{fig:4}), leading to cancelation of $Q_{\text{scat}}^{p+ikT}$. However, this approximate analysis overestimates the scattering suppression that is accurately quantified by $Q_{\text{scat}}^{a_{\text{E}}(1,0)}$. Indeed, the anapole dips in $Q_{\text{scat}}^{a_{\text{E}}(1,0)}$ (red) appear to be much shallower than those in  $Q_{\text{scat}}^{p+ikT}$ (pink). Furthermore, at shorter wavelengths, the lineshapes of the two spectra differ qualitatively. The observed discrepancy can be reduced by introducing higher-order toroidal moments~\cite{gurvitz2019}, correcting the magnitudes of the multipole moments in the regime beyond the point-scatterer approximation. Here, we use a different approach. Instead of adding extra terms to Eqs.~(\ref{eq:pz-small}) and (\ref{eq:Tz}), we replace them with the following generalized expressions (see Eq.~(T2-1) in Ref.~\cite{alaee2018}):
\begin{equation} \label{eq:pz-any}
p_z=\frac{i}{\omega}\int\limits_{-\infty}^{\infty}J_z(\mathbf{r})j_0(kr)d^3\mathbf{r}, 
\end{equation}
\begin{equation} \label{eq:Tz-any}
T_z=\frac{k}{2\omega}\int\limits_{-\infty}^{\infty}\left\{3[\mathbf{r}\cdot\mathbf{J}(\mathbf{r})]z-r^2J_z(\mathbf{r})\right\}\frac{j_2(kr)}{(kr)^2}d^3\mathbf{r}.    
\end{equation}
Writing these expressions in terms of the exact current multipole moments defined in Eq.~(\ref{eq:M_exact}) leads to the following \emph{exact} anapole condition:
\begin{equation} \label{eq:an-cond-any}
p_{z} + \frac{k^{2}}{15}  O_{a_{\text{E}}(1,0)} = 0,
\end{equation}
where
\begin{equation} \label{eq:OaE10}
\begin{aligned}
O_{a_{\text{E}}(1,0)} = \frac{15i}{k}T_z= 3 O_{xxz} + 3 O_{yyz}\qquad \\ + 2 O_{zzz}- O_{zxx} - O_{zyy}.
\end{aligned}
\end{equation}
This condition is the same as the one obtained from Eq.~(\ref{eq:Mapping-for-aE0}) by setting the exact $a_{\text{E}}(1,0)$ to 0. Note that both the general $p_z$ and $O_{a_{\text{E}}(1,0)}$ depend on the choice of the origin of the coordinate system, but the expression $p_z+\frac{k^2}{15}O_{a_{\text{E}}(1,0)}$ is independent of that choice, as it determines the joint scattering contribution of $p_z$ and $O_{a_{\text{E}}(1,0)}$.

\begin{figure}[t]
\centering\includegraphics[width=8.5cm]{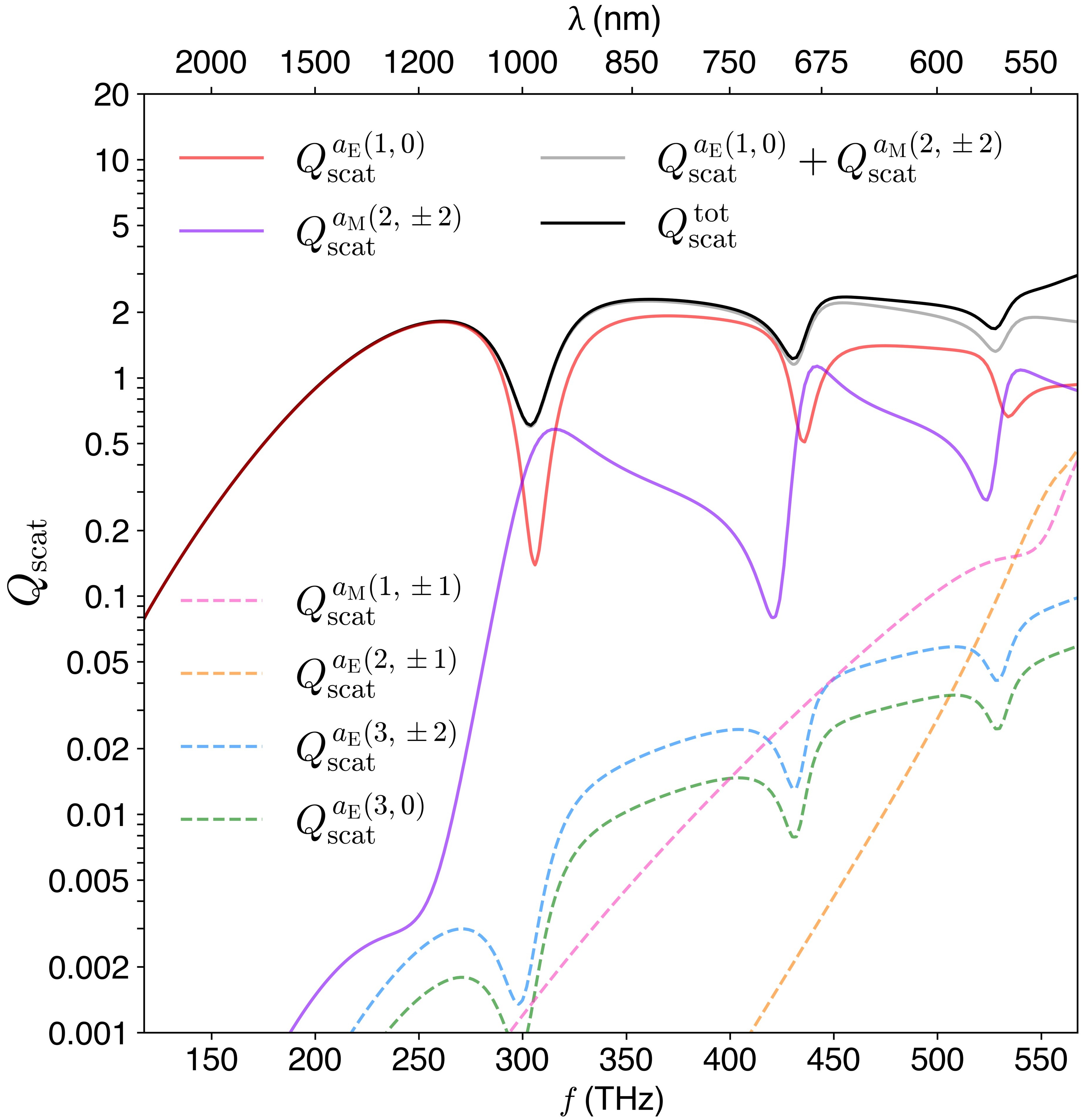}
\caption{\label{fig:5} Selected contributions of classical multipole moments $a_{\text{E/M}}(l,\pm m)$ to the scattering cross section of a silicon nanodisk (the same as in Fig.~\ref{fig:4}). Other contributions are significantly smaller or vanish due to symmetry.}
\end{figure}

\begin{figure}[t]
\centering\includegraphics[width=8.5cm]{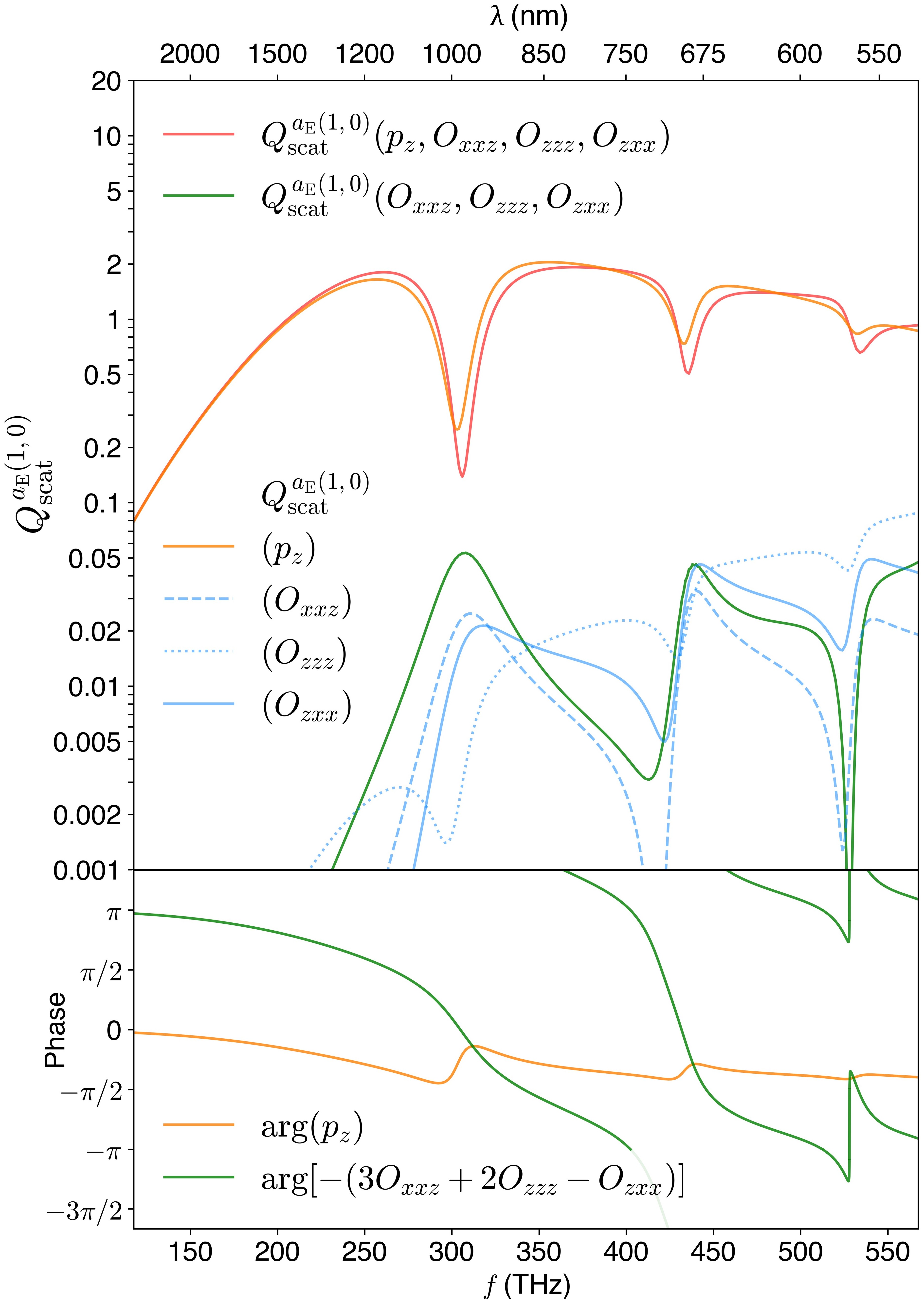}
\caption{\label{fig:6} Contributions of current multipole moments $p_z$, $O_{xxz}$, $O_{zzz}$, and $O_{zxx}$ to the electric-dipole scattering cross section $a_{\text{E}}(1,0)$ of a silicon nanodisk. The lower plot shows the phases of $p_z$ and $-(3O_{xxz}+2O_{zzz}-O_{zxx})$.}
\end{figure}

For point-like scatterers, the exact anapole condition converges to the approximate one given in Eq. (\ref{eq:an-con-small}). Indeed, the spherical Bessel functions in Eqs. (\ref{eq:pz-any}) and (\ref{eq:Tz-any}) can under the small-argument condition be approximated as $j_0(kr)\approx1-(kr)^2/6$ and $j_2(kr)\approx(kr)^2/15$. The dipole moment in Eq.~(\ref{eq:an-cond-any}) can then be written in terms of the point-scatterer moments as
\begin{equation} \label{eq:pz-approx}
p_z\approx p_z^{\text{point}} -\frac{k^2}{3}\left( O_{zxx}^{\text{point}}+O_{zyy}^{\text{point}}+O_{zzz}^{\text{point}}\right),    
\end{equation}
while the exact octupole moments simply become equal to the point-scatterer ones, i.e., $O_{\alpha\beta\gamma}\approx O_{\alpha\beta\gamma}^{\text{point}}$. Inserting these expressions into Eq.~(\ref{eq:an-cond-any}) leads to Eq.~(\ref{eq:an-con-small}).

The obtained exact anapole condition allows us to identify the actual current configurations (i.e., current multipole moments) that are responsible for the formation of anapoles. We test this approach numerically by calculating the contributions of different multipole moments to the scattering cross section of the disk. We start by analyzing the contributions of the classical multipole moments using Eq.~(\ref{eq:scat}). They correspond to orthogonal scattered fields and therefore can be summed together. The non-negligible contributions are shown in Fig.~\ref{fig:5}. The scattering spectrum is essentially dominated by the contributions of $a_{\text{E}}(1,0)$ (electric dipole; red) and $a_{\text{M}}(2,\pm2)$ (magnetic quadrupole; violet). Their sum (grey) overlaps almost perfectly with $Q_{\text{scat}}^{\text{tot}}$ (black) in the studied spectral range. In fact, the parasitic scattering that reduces the visibility of the anapole signatures in the total scattering spectrum mainly results from the magnetic quadrupole excitation represented by $a_{\text{M}}(2,\pm 2)$.

Next, we analyze the contributions of the current multipole moments to the electric-dipole scattering cross section $Q_{\text{scat}}^{a_{\text{E}}(1,0)}$. In general, it involves six moments, but only four of them are significant in the studied case: $p_z$, $O_{xxz}$, $O_{zzz}$, and $O_{zxx}$. Each contribution is evaluated using Eqs.~(\ref{eq:Mapping-for-aE0}) and (\ref{eq:scat}) by setting all other moments to 0. For example, for $O_{xxz}$ we write:
\begin{equation}
    Q_{\text{scat}}^{a_{\text{E}}(1,0)}(O_{xxz})=\frac{3\pi}{k^2S}\left|7\sqrt{2} C_3 O_{xxz}\right|^2.
\end{equation}
The scattered fields produced by the current multipole moments are not orthogonal, and therefore, their scattering contributions are not additive. To calculate a joint contribution of several selected current multipole moments, they must be combined within the corresponding mapping relation. For example, the joint contribution of the moments in question is
\begin{equation}
\begin{aligned}
    Q_{\text{scat}}^{a_{\text{E}}(1,0)}(p_z,O_{xxz},O_{zzz},O_{zxx})=\qquad\qquad\qquad\\\frac{6\pi}{k^2S}\left|C_1 [\,p_z+\frac{k^2}{15}(3O_{xxz}+2O_{zzz}-O_{zxx})]\right|^2.
\end{aligned}
\end{equation}
The above cross section is shown as a red curve in Fig.~\ref{fig:6} and it is essentially identical to $Q_{\text{scat}}^{a_{\text{E}}(1,0)}$ shown in Fig.~\ref{fig:5}, which means that the other terms in Eq.~(\ref{eq:Mapping-for-aE0}) ($O_{yyz}$ and $O_{zyy}$) can indeed be neglected. Interestingly, the isolated contribution of $p_z$ (orange curve in Fig.~\ref{fig:6}) closely follows the curve of $Q_{\text{scat}}^{a_{\text{E}}(1,0)}$, including the anapole dips. The dips coincide with the peaks in the scattering contribution of the current octupoles (green), whereas the total octupole moment $3O_{xxz}+2O_{zzz}-O_{zxx}$, which is responsible for that scattering, is out of phase with $p_z$ (see the lower part of Fig.~\ref{fig:6}). This leads to a slight improvement in the suppression of $Q_{\text{scat}}^{a_{\text{E}}(1,0)}$ (i.e., deeper dips in the red curve compared to those in the orange curve), which can be attributed to a destructive interference analogous to that of $p_z^{\text{point}}$ and $T_z^{\text{point}}$. Despite this, Fig.~\ref{fig:6} clearly shows that the electric-dipole scattering cross section due to the current octupole moments (green) is much smaller than that of $p_z$ (orange). This means that the exact anapole condition defined by Eq.~(\ref{eq:an-cond-any}) is not fulfilled, making the interference between the dipole and octupole contributions inefficient. What we see is that the anapole dips appear directly in $p_z$ and their existence does not depend on any destructive interference with octupole or toroidal moments. The scattering suppression could be improved by modifying the scatterer geometry and/or the illumination conditions to enhance the influence of the octupole/toroidal moments on electric-dipole scattering. In fact, the geometry of a thin disk is fundamentally not suitable for supporting perfectly dark anapole excitations, because it is essentially extended only in two dimensions, which prevents cancelation of scattering simultaneously in all directions. The full cancelation would require also the moments $O_{yyz}$ and $O_{zyy}$ to be excited, which is impossible in such a thin disk geometry.

\begin{figure}[t]
\centering\includegraphics[width=8.5cm]{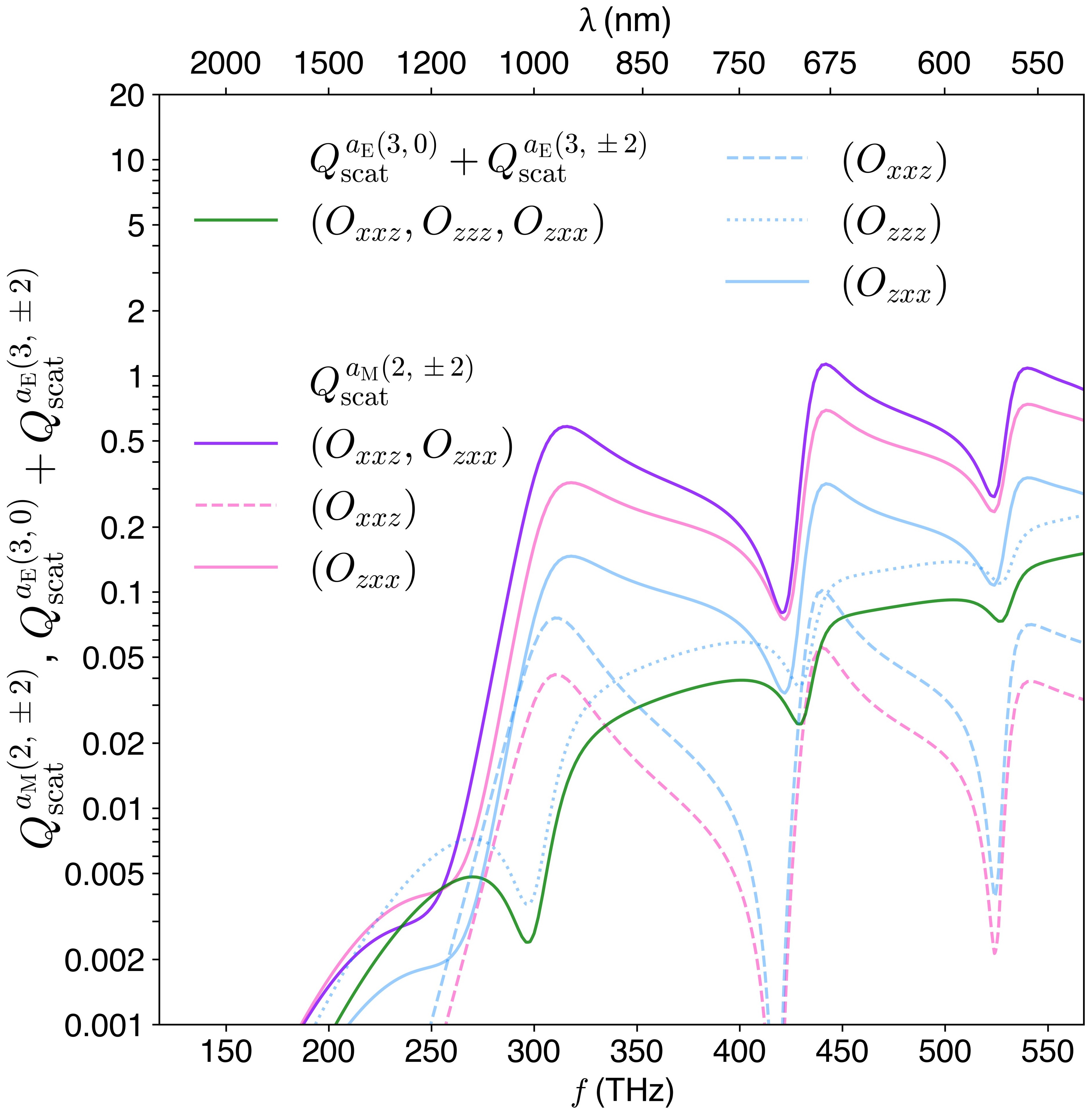}
\caption{\label{fig:7} Contributions of current octupole moments $O_{xxz}$, $O_{zzz}$, and $O_{zxx}$ to the magnetic-quadrupole and electric-octupole scattering cross sections of a silicon nanodisk.}
\end{figure}

Although the contributions of the current octupole moments to the electric-dipole scattering are relatively small, they are responsible for the significant magnetic-quadrupole scattering observed in Fig.~\ref{fig:5} (see Eq.~(\ref{eq:aM22})). Figure~\ref{fig:7} shows the joint contribution of $O_{xxz}$ and $O_{zxx}$ to the magnetic-quadrupole scattering due to $a_{\text{M}}(2,\pm 2)$ (violet curve, identical to $Q_{\text{scat}}^{a_{\text{M}}(2,\pm 2)}$ in Fig.~\ref{fig:5}) with the dominant influence of $O_{zxx}$ (solid pink). In the electric-octupole scattering (green curve), the octupole moments (blue curves) tend to interfere destructively, rendering this type of scattering inefficient.

\begin{figure}[t]
\centering\includegraphics[width=8.5cm]{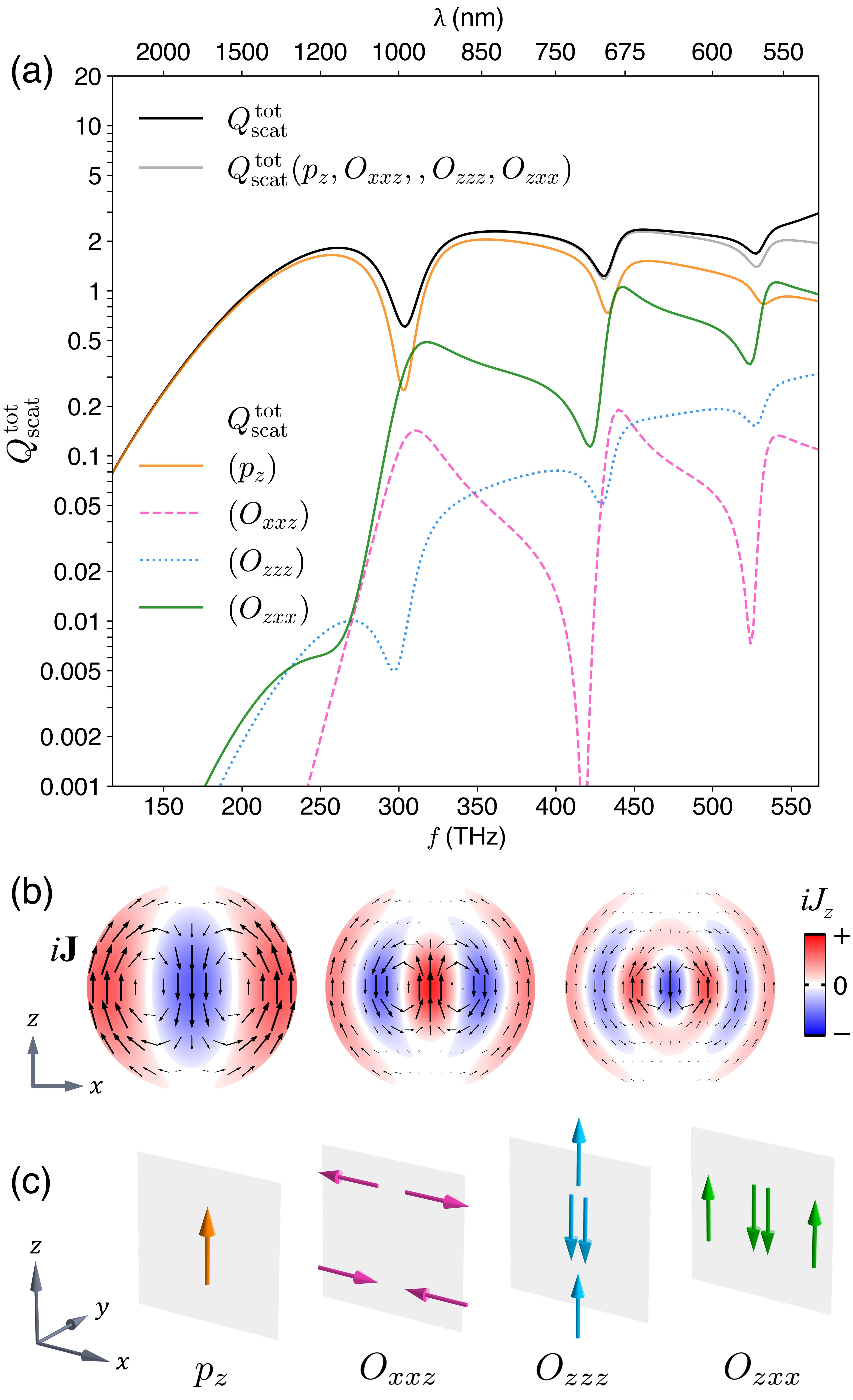}
\caption{\label{fig:8} (a) Contributions of current multipole moments $p_z$ $O_{xxz}$, $O_{zzz}$, and $O_{zxx}$ to the total scattering cross section ($Q_{\text{scat}}^{\text{tot}}$) of a silicon nanodisk. (b) Spatial distribution of $i\mathbf{J}$ (black arrows) and $iJ_z$ (blue-white-red color scale) across the $xz$ slice of the disk at the three subsequent anapole excitations. Both $\mathbf{J}$ and $J_z$ are multiplied by the imaginary unit $i$, which makes them in-phase with the current multipole moments. (c) Illustration of the current multipole moments $p_z$ $O_{xxz}$, $O_{zzz}$, and $O_{zxx}$, which have the most significant contributions to $Q_{\text{scat}}^{\text{tot}}$.}
\end{figure}

Figures~\ref{fig:6} and \ref{fig:7} show that the contributions of the same current multipole to the scattering through different classical multipole moments have identical spectral shapes, differing only by a constant factor (see, for example,  $Q_{\text{scat}}^{a_{\text{E}}(1,0)}(O_{zxx})$, $Q_{\text{scat}}^{a_{\text{M}}(2,\pm 2)}(O_{zxx})$ and $Q_{\text{scat}}^{a_{\text{E}}(3,0)}(O_{zxx})+Q_{\text{scat}}^{a_{\text{E}}(3,\pm 2)}(O_{zxx})$ in Figs.~\ref{fig:6} and \ref{fig:7}). Therefore, it is possible to evaluate the contribution of each current multipole moment to the total scattering cross section. We present such overall contributions in Fig.~\ref{fig:8}(a) for $p_z$, $O_{xxz}$, $O_{zzz}$ and $O_{zxx}$. The fractions of optical power scattered into different radiation patterns (electric dipole, magnetic quadrupole, etc.) by current multipole moments are in fact universal constants, independent of the scatterer and the excitation conditions (provided that the moments are non-zero). Their values for $p_z$, $O_{xxz}$, $O_{zzz}$ and $O_{zxx}$ are summarized in Table~\ref{tab:table2}.

\begin{table}
\caption{\label{tab:table2}%
Fractions of optical power scattered into a specific multipole radiation pattern by current multipole moments $p_z$, $O_{xxz}$, $O_{zzz}$ and $O_{zxx}$.
}
\begin{ruledtabular}
\begin{tabular}{lcccc}
Radiation pattern & $p_z$ & $O_{xxz}$  & $O_{zzz}$  & $O_{zxx}$\\
\colrule\vspace{-2.5mm}\\ 
$a_{\text{E}}(1,0)$ & 1 & 7/40 & 7/25 & 7/160 \\ \vspace{-2.5mm}\\
$a_{\text{M}}(2,\pm 2)$ & 0 & 7/24 & 0 & 21/32 \\ \vspace{-2.5mm}\\
$a_{\text{E}}(3,0)$ & 0 & 1/5 & 18/25 & 9/80  \\ \vspace{-2.5mm}\\
$a_{\text{E}}(3,\pm 2)$ & 0 & 1/3 & 0 & 3/16  \\ \vspace{-2.5mm}\\
\end{tabular}
\end{ruledtabular}
\end{table}

The spatial distributions of $\mathbf{J}$ in Fig.~\ref{fig:8}(b) reveal the presence of multipole excitations beyond octupoles, especially in the higher-order anapoles. However, the scattering contributions of higher-order multipoles are found to be negligible, which means that they play no essential role in the formation of anapole dips in $Q_{\text{scat}}^{\text{tot}}$. In summary, our analysis shows that the scattering properties of the studied system are almost entirely determined by the current multipoles illustrated in Fig.~\ref{fig:8}(c). An interesting conclusion is that the current octupole moments (which are the building blocks of the toroidal electric dipole moment) do not contribute significantly to the formation of anapoles in disk-shaped scatterers, but they in fact produce parasitic magnetic-quadrupole scattering that makes the anapole dips in $Q_{\text{scat}}^{\text{tot}}$ much shallower.

The exact anapole condition in Eq.~\ref{eq:an-cond-any} can be generalized to multipole scattering of an arbitrary order $l$ by setting the relevant classical multipole expansion coefficients $a_{\text{E/M}}(l,m)$ to 0. For electric-multipole scattering, the exact anapole condition can be written as
\begin{equation}\label{eq:an-con-aEany}
M^{(l)}_{a_{\text{E}}(l,m)} +c_l M^{(l+2)}_{a_{\text{E}}(l,m)}=0,
\end{equation}
where $M^{(l)}_{a_{\text{E}}(l,m)}$ and $M^{(l+2)}_{a_{\text{E}}(l,m)}$ are the sums of the current multipole moments of orders $l$ and $l+2$, respectively, contributing to the $l$-order electric-multipole scattering described by coefficient $a_{\text{E}}(l,m)$, while $c_l$ is the relevant ratio of the coefficients in the corresponding mapping relation (involving a ratio between $C_l$ and $C_{l+2}$). Equation~(\ref{eq:an-con-aEany}) resembles the conventional anapole condition, with $M^{(l+2)}_{a_{\text{E}}(l,m)}$ playing the role of a toroidal multipole moment, i.e., the moment with a more complex current configuration that produces the same radiation pattern as the multipole moment $M^{(l)}_{a_{\text{E}}(l,m)}$. This analogy does not hold for magnetic-multipole scattering, in which case the anapole condition imposes a relation only between current multipole moments of the same order ($l+1$). For example, in the case of magnetic-dipole scattering associated with $a_{\text{M}}(1,0)$, the anapole condition is 
\begin{equation}
Q_{xy}-Q_{yx} = 0, 
\end{equation}
and it involves only current quadrupoles (see Eq.~(\ref{eq:Mapping-for-aM0})). Note, however, that fulfilling this condition inherently leads to a non-zero contribution to the electric-quadrupole scattering associated with $a_{\text{E}}(2,\pm2)$, which contains the sum $Q_{xy}+Q_{yx}$. This situation is similar to the studied example of a nanodisk, in which current octupoles produced significant magnetic-quadrupole scattering despite relatively weak electric-dipole and electric-octupole scattering.

In the above examples, we have mainly focused on the contributions of the current multipole moments to the scattering cross section. However, these moments can also be used to assess the magnitude of the corresponding current density and the resulting local field enhancement~\cite{sehrawat2024_,sehrawat2024__}. In Appendix B, we establish a direct connection between the current multipoles and the current density at the surface of a scatterer. Rigorous analysis of this approach requires separate attention and, therefore, we leave it for a future work.

\section*{Conclusions}

In summary, we have introduced an expression for the general and exact current multipole tensor (see Eq. (\ref{eq:M_exact})), enabling rigorous analysis of localized current configurations in electromagnetic scatterers. The expressions for current multipole moments remain very simple also for large multipole orders (see Table \ref{tab:table1}). Unlike the classical field-based multipoles, current multipoles are not divided into electric and magnetic types, which reflects their universal nature. Despite their simple and intuitive form, they can be used to quantitatively determine the multipole contributions to the scattering and extinction cross sections for scatterers of arbitrary sizes and shapes. We have verified our expression numerically for wavelength-scale spherical scatterers supporting higher-order multipole excitations and found excellent agreement with the Mie theory. Furthermore, we demonstrated the usefulness of our approach in describing nonradiating current configurations in the well-known example of a dielectric nanodisk supporting anapole excitations. In particular, our analysis reveals the true nature of the so-called toroidal dipole moment, showing that it is composed of current octupole moments, being responsible for parasitic magnetic-quadrupole scattering in a thin disk geometry. Finally, current multipole expansion allowed us to derive the exact anapole condition and to generalize it to multipole scattering of an arbitrary order, opening new avenues in the design of anapoles and other nonradiating states of light. Overall, our approach simplifies the analysis of multipole excitations and the design of arbitrarily sized individual and arrayed electromagnetic scatterers, which can be useful in many areas of optics and photonics, including optical antennas and metamaterials.

The data that support the findings of this article are openly available~\cite{kolkowski2025}.

The authors acknowledge the support of the Research Council of Finland (Grants No. 347449, 353758 and 368485). R. K. has received funding from the European Union’s Horizon Europe programme for research and innovation under the Marie Sk\l{}odowska-Curie Grant Agreement No. 101060306 (NExIA). For computational resources, the authors acknowledge the Aalto University School of Science “Science-IT” project and CSC – IT Center for Science, Finland.

\appendix

\section{Derivation of Equation (\ref{eq:M_exact}).}

To expand the scattering current density \textbf{J}(\textbf{r}) defined in Eq.~(\ref{eq:J}), one can use the identity
\begin{equation} \label{eq:JJ}
\textbf{J}(\textbf{r})=\int\limits_{-\infty}^{\infty} \textbf{J}(\textbf{r}')\delta(\textbf{r}'-\textbf{r})d^3\textbf{r}',
\end{equation}
where $\delta(\textbf{r}'-\textbf{r})$ is the three-dimensional Dirac delta function. Expanding the delta function into a Taylor series about $\textbf{r}'=0$, we obtain
\begin{eqnarray}\label{eq:J-expanded}
\textbf{J}(\textbf{r})=\,i\omega\sum_{l=1}^\infty\sum_{\hat{\textbf{v}}=\hat{\textbf{x}},\hat{\textbf{y}},\hat{\textbf{z}}}\sum_{a=0}^{l-1}\sum_{b=0}^{l-a-1}M^{(l)}(\hat{\textbf{v}},a,b)\hat{\textbf{v}}\qquad\qquad \nonumber \\
\times\frac{(-1)^l(l-1)!}{a!b!(l-a-b-1)!}\frac{\mathrm{d}^a}{\mathrm{d}x^a}\frac{\mathrm{d}^b}{\mathrm{d}y^b}\frac{\mathrm{d}^{l-a-b-1}}{\mathrm{d}z^{l-a-b-1}}\delta(\textbf{r}),\qquad
\end{eqnarray}
where $\hat{\textbf{x}}$, $\hat{\textbf{y}}$, and $\hat{\textbf{z}}$ are the unit vectors along the corresponding Cartesian coordinate axes and $M^{(l)}(\hat{\textbf{v}},a,b)$ are the elements of the multipole tensor $\boldsymbol{M}^{(l)}$ given in Eq.~(\ref{eq:M_point}). These elements can be written as
\begin{equation} \label{eq:M^l_ab}
M^{(l)}(\hat{\textbf{v}},a,b) = \frac{i}{(l-1)!\,\omega}\int\limits_{-\infty}^{\infty} J_{v}(\textbf{r}) x^{a} y^{b} z^{l-a-b-1}d^3\textbf{r},    
\end{equation}
where $J_{v}(\textbf{r})$ is the \textit{v}-component of vector $\textbf{J}(\textbf{r})$. Equation~(\ref{eq:M^l_ab}) coincides with Eq.~(\ref{eq:M_point}) and is valid only if the size of the scatterer, $L$, satisfies the condition $kL\ll1$. Using the identity
\begin{equation}
    \frac{\mathrm{d}^n}{\mathrm{d}x^n}\delta(\textbf{x})=\frac{n!(-1)^n}{x^n}\delta(\textbf{x}),
\end{equation}
one can rewrite Eq.~(\ref{eq:J-expanded}) in the form
\begin{eqnarray}\label{eq:J-expanded-2}
\textbf{J}(\textbf{r})=\,-i\omega\sum_{l=1}^\infty\sum_{\hat{\textbf{v}}=\hat{\textbf{x}},\hat{\textbf{y}},\hat{\textbf{z}}}\sum_{a=0}^{l-1}\sum_{b=0}^{l-a-1}M^{(l)}(\hat{\textbf{v}},a,b)\hat{\textbf{v}}\qquad\qquad \nonumber \\
\times\frac{(l-1)!}{x^ay^bz^{l-a-b-1}}\delta(\textbf{r}),\qquad
\end{eqnarray}
which is in agreement with Eq. (\ref{eq:M^l_ab}). The right-hand side of this expression is a sum of the current densities associated with the multipole moments. These current densities are
\begin{equation} \label{eq:J_lvab-point}
J^{(l)}_{\hat{\textbf{v}},a,b}(\textbf{r})=-i\omega M^{(l)}(\hat{\textbf{v}},a,b)\frac{(l-1)!}{x^ay^bz^{l-a-b-1}}\delta(\textbf{r}).
\end{equation}

A similar multipole expansion can be written for the vector potential \textbf{A}(\textbf{r}). Note that, in contrast to the electric and magnetic fields, the field of the vector potential is produced by electromagnetic anapoles as well. In the Lorentz gauge, the vector potential generated by a localized scattering current density obeys the Helmholtz equation: 
\begin{equation} \label{Helmholtz-A}
    (\nabla^2+k^2)\textbf{A}(\textbf{r})=-\mu\textbf{J}(\textbf{r}),
\end{equation}
where $k$ and $\mu$ are the wavenumber and magnetic permeability in the surrounding medium. The solution of this equation can be written as the following integral:
\begin{equation} \label{eq:A-J}
    \textbf{A}(\textbf{r})=\mu\int\limits_{-\infty}^{\infty}\textbf{J}(\textbf{r}')G(\textbf{r}'-\textbf{r})d^3\textbf{r}',
\end{equation}
where $G(\textbf{r}'-\textbf{r})=e^{ik|\textbf{r}'-\textbf{r}|}/4\pi|\textbf{r}'-\textbf{r}|$ is the Green's function for a point source at a coordinate $\textbf{r}'$ that is proportional to the spherical Hankel function of the first kind and zeroth order $h_0^{(1)}(k|\textbf{r}'-\textbf{r}|)$. Using the identity $G(\textbf{r}'-\textbf{r})=(ik/4\pi)h_0^{(1)}(k|\textbf{r}'-\textbf{r}|)$ and expanding the Green's function in Eq.~(\ref{eq:A-J}) into a Taylor series about $\textbf{r}'=0$, we obtain
\begin{eqnarray}
\textbf{A}(\textbf{r})=\,\frac{k^3}{4\pi\epsilon\omega}\sum_{l=1}^\infty\sum_{\hat{\textbf{v}}=\hat{\textbf{x}},\hat{\textbf{y}},\hat{\textbf{z}}}\sum_{a=0}^{l-1}\sum_{b=0}^{l-a-1} M^{(l)}(\hat{\textbf{v}},a,b)\hat{\textbf{v}}\qquad\nonumber \\
\times\frac{(-1)^{l-1}(l-1)}{a!b![l-(a+b+1)]}\frac{\mathrm{d}^a}{\mathrm{d}x^a}\frac{\mathrm{d}^b}{\mathrm{d}y^b}\frac{\mathrm{d}^{l-a-b-1}}{\mathrm{d}z^{l-a-b-1}}h_0^{(1)}(kr) ,\qquad
\end{eqnarray}
which is valid for $kL \ll 1$ and is the same as Eq.~(32) in Ref.~\cite{grahn2012}.

Equation (\ref{eq:A-J}) is general and can be used to find the exact expressions for the current multipole moments by expanding $G(\textbf{r}'-\textbf{r})$ in terms of spherical harmonics $Y_{i,j}(\theta,\phi)$. This leads to the following result~\cite{jackson2012}: 
\begin{eqnarray}\label{A-in-sph-harm}
\textbf{A}(\textbf{r})=&&\,i\mu\,\sum_{l=1}^\infty\,\sum_{\hat{\textbf{v}}=\hat{\textbf{x}},\hat{\textbf{y}},\hat{\textbf{z}}} \,\sum_{m=-l+1}^{l-1}k^lh_{l-1}^{(1)}(kr)\nonumber\\&&\times Y_{l-1,m}(\theta,\phi)\tilde{M}^{(l)}(\hat{\textbf{v}},m)\hat{\textbf{v}},
\end{eqnarray}
where $r$, $\theta$, and $\phi$ are the spherical coordinates, and the coefficients $\tilde{M}^{(l)}(\hat{\textbf{v}},m)$ are given by
\begin{equation}
\tilde{M}^{(l)}(\hat{\textbf{v}},m)=\int\limits_{-\infty}^{\infty} J_v(\textbf{r})\frac{j_{l-1}(kr)}{k^{l-1}}Y_{l-1,m}^*(\theta,\phi)d^3\textbf{r}. \end{equation}
These coefficients are the current multipole moments in the expansion of $\textbf{A}(\textbf{r})$. They possess mutually orthogonal current distributions and can be used as a basis for expansion of arbitrary localized current densities. Let us find several lowest-order moments $\tilde{M}^{(l)}(\hat{\textbf{v}},m)$ with the spherical harmonics in their expressions written in Cartesian coordinates:
\begin{eqnarray}
&&\tilde{M}^{(1)}(\hat{\textbf{v}},0)=\frac{1}{2}\sqrt{\frac{1}{\pi}}\int\limits_{-\infty}^{\infty} J_v(\textbf{r})j_{0}(kr)d^3\textbf{r}, \\
&&\tilde{M}^{(2)}(\hat{\textbf{v}},0)=\frac{1}{2}\sqrt{\frac{3}{\pi}}\int\limits_{-\infty}^{\infty} J_v(\textbf{r})z\frac{j_{1}(kr)}{kr}d^3\textbf{r}, \nonumber\\
&&\tilde{M}^{(2)}(\hat{\textbf{v}},\pm 1)\!=\!\mp\frac{1}{2}\sqrt{\frac{3}{2\pi}}\int\limits_{-\infty}^{\infty} J_v(\textbf{r})(x \pm iy)\frac{j_{1}(kr)}{kr}d^3\textbf{r}, \nonumber\\
&&\tilde{M}^{(3)}(\hat{\textbf{v}},0)=\frac{1}{4}\sqrt{\frac{5}{\pi}}\int\limits_{-\infty}^{\infty} J_v(\textbf{r})(3z^2-r^2)\frac{j_{2}(kr)}{(kr)^2}d^3\textbf{r}, \nonumber\\
&&\tilde{M}^{(3)}(\hat{\textbf{v}},\pm 1)=\mp \frac{1}{2}\sqrt{\frac{15}{2\pi}}\int\limits_{-\infty}^{\infty} J_v(\textbf{r})(xz\pm iyz)\frac{j_{2}(kr)}{(kr)^2}d^3\textbf{r} \nonumber.
\end{eqnarray}
It can be seen that each moment $\tilde{M}^{(l)}(\hat{\textbf{v}},m)$ is a simple linear combination of the following normalized moments:
\begin{eqnarray}
M^{(l)}(\hat{\textbf{v}},a,b) =&& \,\frac{i}{\omega}\frac{(2l-1)!!}{(l-1)!}\int\limits_{-\infty}^{\infty} J_{v}(\textbf{r}) \label{M^l}\nonumber \\
&&\times x^{a} y^{b} z^{l-a-b-1}\frac{j_{l-1}(kr)}{(kr)^{l-1}}d^3\textbf{r}\label{eq:M^l_ab_exact}.\qquad  
\end{eqnarray}
These moments form the general scattering-current multipole tensor defined in Eq.~(\ref{eq:M_exact}). In the limit of $kr\ll1$, $j_{l-1}(kr)/(kr)^{l-1}$ approaches $1/(2l-1)!!$~\cite{jackson2012}, and Eq.~(\ref{eq:M^l_ab_exact}) converges to Eq.~(\ref{eq:M^l_ab}). Correspondingly, Eq.~(\ref{eq:M_exact}) converges to Eq.~(\ref{eq:M_point}).

\section{Current density associated with current multipoles.}

For a point-like scatterer, Eq.~(\ref{eq:J_lvab-point}) connects the current multipole moments with the associated scattering current densities. For a realistic scatterer that is much smaller than the wavelength, the current density in a dipole excitation is nearly uniform. Hence, the volume-averaged current density can be equated to the current density itself, and we have
\begin{equation}
J^{(1)}_{\hat{\textbf{v}},0,0}(\textbf{r})=-i\omega p_v\delta(\textbf{r})=-i\omega p_v\frac{1}{V},   
\end{equation}
where $V$ is the volume of the scatterer. Then, to keep the relative strengths of the multipolar current densities unchanged, we replace $\delta(\textbf{r})$ with $1/V$ also for other orders $l$ in Eq.~(\ref{eq:J_lvab-point}). Furthermore, in the considered model, the current density is required to produce the actual field only outside the scatterer, and the field and current density inside the scatterer can be arbitrary. Therefore, we expect the calculated current density to correspond to the realistic current density only on the surface of the scatterer. If the radial sizes of the scatterer along the $x$, $y$, and $z$ directions are $X, Y$, and $Z$, respectively (such that the volume of the scatterer is $V\approx 8XYZ$), the approximate expression for the multipole current density on the surface can be written as
\begin{equation} \label{eq:J_lvab-real}
J^{(l)}(\hat{\textbf{v}},a,b)\approx -\frac{i}{8}\frac{\omega(l-1)!}{X^{a+1}Y^{b+1}Z^{l-a-b}}M^{(l)}(\hat{\textbf{v}},a,b).
\end{equation}
This equation allows assessing the surface current density and the corresponding near-field enhancement associated with a given multipole moment. The phase of $J^{(l)}(\hat{\textbf{v}},a,b)$ is equal to that of the outermost current element in the given multipole at positive $x$, $y$, and $z$.

The generalized surface current density associated with each multipole moment can be written in the same form as Eq.~(\ref{eq:J_lvab-real}), but with $V=8XYZ$ being replaced with 
\begin{equation}
    I=(2l-1)!!\int\limits_V \frac{j_{l-1}(kr)}{(kr)^{l-1}}d^3\textbf{r}.
\end{equation}
Here, the integration is performed over the volume of the scatterer. This gives the following result:
\begin{equation} \label{eq:J_lvab-gen}
J^{(l)}(\hat{\textbf{v}},a,b)\approx -\frac{i}{I}\frac{\omega(l-1)!}{X^{a}Y^{b}Z^{l-a-b-1}}M^{(l)}(\hat{\textbf{v}},a,b).
\end{equation}
If the scatterer is much smaller than the wavelength, $I$ is equal to $V$, which makes Eq.~(\ref{eq:J_lvab-gen}) identical to Eq.~(\ref{eq:J_lvab-real}).

\section{Classical multipole moments as linear superpositions of current multipole moments.}

Exact expressions for $a_\text{E}(l,m)$ and $a_\text{M}(l,m)$ with $l\geq 2$ in terms of current multipole moments up to triacontadipoles are given in Eqs.~(\ref{eq:aE20})-(\ref{eq:aM44}), supplementing the expressions for the electric and magnetic dipoles given in Eqs.~(\ref{eq:Mapping-for-aE0})-(\ref{eq:Mapping-for-aM1}). The factors $C_1$-$C_3$ in the expressions are defined below Eq.~(\ref{eq:Mapping-for-aM1}) and the two new factors are  $C_4=- k^{6}/(350 \pi \epsilon E_{0} )$ and $C_5 = - i k^{7}/(4410 \pi \epsilon E_{0} )$. The expressions for any order $l$ can be derived automatically, following instructions provided in Ref.~\cite{kolkowski2025}.

\begin{widetext}
\begin{eqnarray}\label{eq:aE20}
a_\text{E}(2,0) =\nonumber&&\,-\sqrt{6} \, C_{2}\, \left(\,Q_{xx} + Q_{yy} - 2 Q_{zz}\,\right) \nonumber\\
&&-\sqrt{6}\,C_{4} \,(\,H_{xxxx} + H_{xxyy} - 4 H_{xxzz} + H_{yxxy}+ H_{yyyy}  \nonumber\\
&&\qquad\qquad- 4 H_{yyzz}+ 3 H_{zxxz} + 3 H_{zyyz} - 2 H_{zzzz}\,)\end{eqnarray}

\begin{eqnarray}
a_\text{E}(2,\pm1) =\nonumber&&\, \mp \,3 C_{2} \left(\,Q_{xz} + Q_{zx}\right) +\,3i C_{2} \left(\,Q_{yz} + Q_{zy}\,\right) \\ &&\mp \,2 C_{4} \, \left(\,4 H_{xxxz} - H_{xyyz} - H_{xzzz} + 5 H_{yxyz} - H_{zxxx} - H_{zxyy} + 4 H_{zxzz}\,\right)\nonumber \\&&-\, 2i C_{4} \left(\,4 H_{yyyz} - H_{yxxz}  - H_{yzzz} + 5 H_{xxyz} - H_{zxxy} - H_{zyyy} + 4 H_{zyzz}\,\right)\end{eqnarray}

\begin{eqnarray}
a_\text{E}(2,\pm2) =\nonumber&&\, 3 C_{2} \left[\, Q_{xx} - Q_{yy} \mp i \left(Q_{xy} + Q_{yx}\,\right)\,\right]\nonumber \\&&+\, C_{4} \, \left(\,3 H_{xxxx} - 7 H_{xxyy} - 2 H_{xxzz} + 7 H_{yxxy} - 3 H_{yyyy} + 2 H_{yyzz} + 5 H_{zxxz} - 5 H_{zyyz}\,\right)\nonumber \\&&\mp\, 2 i C_{4} \left(\, 4 H_{xxxy} - H_{xyyy} - H_{xyzz} - H_{yxxx} + 4 H_{yxyy} - H_{yxzz} + 5 H_{zxyz}\,\right)\,
\end{eqnarray}

\begin{eqnarray}\label{eq:aE30}
a_\text{E}(3,0) =\nonumber&&\, 2 \sqrt{3}\,C_{3} \, \left(\,2 O_{xxz} + 2 O_{yyz} + O_{zxx} + O_{zyy} - 2 O_{zzz}\,\right) \nonumber\\&&
+ \,2 \sqrt{3}\, C_{5} \, (\,15 T_{xxxxz} + 15 T_{xxyyz} - 20 T_{xxzzz} + 15 T_{yxxyz} + 15 T_{yyyyz} - 20 T_{yyzzz} \nonumber\\&&\qquad\qquad
-\, 3 T_{zxxxx} - 6 T_{zxxyy} + 24 T_{zxxzz} - 3 T_{zyyyy} + 24 T_{zyyzz} - 8 T_{zzzzz}\,)
\end{eqnarray}

\begin{eqnarray}\label{eq:aE31}
a_\text{E}(3,\pm1) =&&\, \mp C_{3}  \left(\,3 O_{xxx} + O_{xyy} - 4 O_{xzz} + 2 O_{yxy} - 8 O_{zxz}\,\right)  \nonumber\\&&+ iC_{3} \left(\, 3 O_{yyy} + O_{yxx}  - 4 O_{yzz} + 2 O_{xxy} - 8 O_{zyz}\,\right) \nonumber\\&&\mp\, 3 C_{5} \, (\,4 T_{xxxxx} + 3 T_{xxxyy} + 5 T_{yxyyy} - T_{xyyyy} + 4 T_{xzzzz} + 3 T_{xyyzz} + 5 T_{yxxxy}\nonumber\\&&\qquad\quad - 27 T_{xxxzz}  - 30 T_{yxyzz} + 15 T_{zxxxz} + 15 T_{zxyyz} - 20 T_{zxzzz}\,) \nonumber\\&&+3i C_{5} \, (\,4 T_{yyyyy} + 3 T_{yxxyy} + 5 T_{xxyyy} - T_{yxxxx} + 4 T_{yzzzz} + 3 T_{yxxzz} + 5 T_{xxxxy}  \nonumber\\&&  \qquad\quad- 27 T_{yyyzz} - 30 T_{xxyzz} + 15 T_{zxxyz} + 15 T_{zyyyz} - 20 T_{zyzzz}\,)
\end{eqnarray}

\begin{eqnarray}\label{eq:aE32}
a_\text{E}(3,\pm2) =&& \, \sqrt{10}\,C_{3} \,[\,- 2 O_{xxz} + 2 O_{yyz} - O_{zxx} + O_{zyy} \pm 2 i \left(\,O_{xyz} + O_{yxz} + O_{zxy}\right)\,] \nonumber\\&&- 3\sqrt{10}\, C_{5} \, (\,5 T_{xxxxz} - 9 T_{xxyyz} - 2 T_{xxzzz} + 9 T_{yxxyz} - 5 T_{yyyyz} + 2 T_{yyzzz} \nonumber\\&&\qquad\qquad\quad - T_{zxxxx} + 6 T_{zxxzz} + T_{zyyyy} - 6 T_{zyyzz}\,) \nonumber\\&&\pm\, 6 \sqrt{10}\,i\, C_{5} \,(\, 6 T_{xxxyz} - T_{xyyyz} - T_{xyzzz} - T_{yxxxz} + 6 T_{yxyyz} - T_{yxzzz} \nonumber\\&&\qquad\qquad\quad- T_{zxxxy} - T_{zxyyy} + 6 T_{zxyzz}\,)
\end{eqnarray}

\begin{eqnarray}\label{eq:aE33}
a_\text{E}(3,\pm3) =&&\,   \sqrt{15}\,C_{3} \left[\,\pm\left(\, O_{xxx} - O_{xyy} - 2 O_{yxy}\,\right) - i \left(\,2 O_{xxy} + O_{yxx} - O_{yyy}\,\right)\,\right] \nonumber\\&&\pm \sqrt{15}\,C_{5} \, (\,4 T_{xxxxx} - 21 T_{xxxyy} - 15 T_{yxyyy} + 13 T_{yxxxy} + 3 T_{xyyyy} + 3 T_{xyyzz} \nonumber\\&&\qquad\qquad\quad - 3 T_{xxxzz} + 6 T_{yxyzz} + 7 T_{zxxxz} - 21 T_{zxyyz}\,) \nonumber\\&&+ \sqrt{15}\,i\,C_{5} \,(\,4 T_{yyyyy} - 21 T_{yxxyy}- 15 T_{xxxxy} + 13 T_{xxyyy}  + 3 T_{yxxxx}  + 3 T_{yxxzz}\nonumber\\&&\qquad\qquad\quad - 3 T_{yyyzz} + 6 T_{xxyzz} + 7 T_{zyyyz} - 21 T_{zxxyz}\,)\,]
\end{eqnarray}

\begin{eqnarray}
a_\text{M}(2,0) = 7 \sqrt{6} \,i\, C_{3} \left(\,O_{xyz} - O_{yxz}\,\right)
\end{eqnarray}

\begin{eqnarray}
a_\text{M}(2,\pm 1) = - 7 C_{3}\, [\,O_{xyy} - O_{xzz} - O_{yxy} + O_{zxz} \pm i \left(\,O_{xxy} - O_{yxx} + O_{yzz} - O_{zyz}\,\right)\,]
\end{eqnarray}

\begin{eqnarray}\label{eq:aM22}
a_\text{M}(2,\pm 2) = 7 C_{3} \left[\,\mp \left(\,O_{xxz} - O_{yyz} - O_{zxx} + O_{zyy} \,\right)+ i \left(\,O_{xyz} + O_{yxz} - 2 O_{zxy}\,\right)\,\right]
\end{eqnarray}

\begin{eqnarray}
a_\text{M}(3,0) = \frac{5 }{2}\sqrt{3} \,i\, C_{4} \,\left(\,- H_{xxxy} - H_{xyyy} + 4 H_{xyzz} + H_{yxxx} + H_{yxyy} - 4 H_{yxzz}\,\right)
\end{eqnarray}

\begin{eqnarray}
a_\text{M}(3, \pm 1) =&&\, - \frac{5 }{4}C_{4} [\,H_{xxxz} + 11 H_{xyyz} - 4 H_{xzzz} - 10 H_{yxyz} - H_{zxxx} - H_{zxyy} + 4 H_{zxzz} \nonumber \\&&\qquad\quad \mp \, i \left(\,H_{yyyz} + 11 H_{yxxz} - 4 H_{yzzz} - 10 H_{xxyz} - H_{zyyy} - H_{zxxy}  + 4 H_{zyzz}\,\right)\,]
\end{eqnarray}

\begin{eqnarray}
a_\text{M}(3, \pm 2) =&&\, \frac{5 }{4}\sqrt{10}\, C_{4} \, [\,\pm\left(\,2 H_{xxyy} - 2 H_{xxzz} - 2 H_{yxxy} + 2 H_{yyzz} + 2 H_{zxxz} - 2 H_{zyyz}\,\right) \nonumber \\&&\qquad\qquad\quad+ i \left(H_{xxxy} - H_{xyyy} + 2 H_{xyzz} - H_{yxxx} + H_{yxyy} + 2 H_{yxzz} - 4 H_{zxyz}\,\right)\,]
\end{eqnarray}

\begin{eqnarray}
a_\text{M}(3, \pm 3) =&&\, \frac{5}{4} \sqrt{15} \, C_{4}\, [\,H_{xxxz} - H_{xyyz} - 2 H_{yxyz} - H_{zxxx} + 3 H_{zxyy} \nonumber \\&&\qquad\quad\quad\pm\, i \left(\,H_{yyyz} - H_{yxxz} -2 H_{xxyz} - H_{zyyy} + 3 H_{zxxy} \,\right)\,]
\end{eqnarray}

\begin{eqnarray}\label{eq:aM40}
a_\text{M}(4, 0) = 14 \sqrt{5} \, i \, C_{5} \, \left(\,3 T_{xxxyz} + 3 T_{xyyyz} - 4 T_{xyzzz} - 3 T_{yxxxz} - 3 T_{yxyyz} + 4 T_{yxzzz}\,\right)
\end{eqnarray}

\begin{eqnarray}\label{eq:aM41}
a_\text{M}(4,\pm 1) =&&\, - 7 C_{5} \, [\,3 T_{xxxyy} - 3 T_{xxxzz} + 3 T_{xyyyy} - 21 T_{xyyzz} + 4 T_{xzzzz} - 3 T_{yxxxy} \nonumber \\&&\qquad\quad- 3 T_{yxyyy} + 18 T_{yxyzz} + 3 T_{zxxxz} + 3 T_{zxyyz} - 4 T_{zxzzz} \nonumber \\&&\qquad\quad \mp\, i\, (\, 3 T_{yxxyy} - 3 T_{xxxxy} + 3 T_{yxxxx}  - 21 T_{yxxzz} + 4 T_{yzzzz} - 3 T_{xxyyy} \nonumber \\&&\qquad\qquad- 3 T_{yyyzz} + 18 T_{xxyzz} + 3 T_{zxxyz} + 3 T_{zyyyz} - 4 T_{zyzzz}\,)\,]
\end{eqnarray}

\begin{eqnarray}\label{eq:aM42}
a_\text{M}(4,\pm 2) =&&\, - 7 \sqrt{2} C_{5}\, [\,\pm \, (\, T_{xxxxz} + 15 T_{xxyyz} - 6 T_{xxzzz} - 15 T_{yxxyz} - T_{yyyyz} + 6 T_{yyzzz}  \nonumber \\&&\qquad\qquad\quad - T_{zxxxx} + 6 T_{zxxzz} + T_{zyyyy} - 6 T_{zyyzz}\,) \nonumber \\&&\qquad\qquad+ 2 i \,(\,3 T_{xxxyz} - 4 T_{xyyyz} + 3 T_{xyzzz} - 4 T_{yxxxz} + 3 T_{yxyyz} \nonumber \\&&\qquad\qquad\quad + 3 T_{yxzzz} + T_{zxxxy} + T_{zxyyy} - 6 T_{zxyzz}\,)\,]
\end{eqnarray}

\begin{eqnarray}\label{eq:aM43}
a_\text{M}(4,\pm 3) =&&\, 7 \sqrt{7} C_{5} \, [\,3 T_{xxxyy} - 3 T_{yxxxy} - T_{xyyyy} + T_{yxyyy} + 3 T_{xyyzz} 
\nonumber \\&&\qquad\qquad  - 3 T_{xxxzz} + 6 T_{yxyzz} + 3 T_{zxxxz} - 9 T_{zxyyz} 
\nonumber \\&&\qquad\quad \pm \, i\, (\,3 T_{yxxyy} - 3 T_{xxyyy} - T_{yxxxx} + T_{xxxxy} + 3 T_{yxxzz} \nonumber \\&&\qquad\qquad - 3 T_{yyyzz}  + 6 T_{xxyzz} + 3 T_{zyyyz} - 9 T_{zxxyz} \,)\,]
\end{eqnarray}

\begin{eqnarray}\label{eq:aM44}
a_\text{M}(4,\pm4) =&&\, - 7 \sqrt{14} C_{5} \,[\,\mp \left(\,T_{xxxxz} - 3 T_{xxyyz} - 3 T_{yxxyz} + T_{yyyyz} - T_{zxxxx} + 6 T_{zxxyy} - T_{zyyyy}\,\right)\nonumber \\&&\qquad\qquad\quad
+ i\, (\,3 T_{xxxyz} - T_{xyyyz} + T_{yxxxz} - 3 T_{yxyyz} - 4 T_{zxxxy} + 4 T_{zxyyy}\,)\,]
\end{eqnarray}

\end{widetext}

\bibliography{lib}

\end{document}